# Automated Modal Parameter Estimation Using Correlation Analysis and Bootstrap Sampling

Vahid Yaghoubi, Majid K. Vakilzadeh[1], Thomas J.S. Abrahamsson

Department of Applied Mechanics, Chalmers University of Technology, Gothenburg, Sweden

**ABSTRACT**

The estimation of modal parameters from a set of noisy measured data is a highly judgmental task, with user expertise playing a significant role in distinguishing between estimated physical and noise modes of a test-piece. Various methods have been developed to automate this procedure. The common approach is to identify models with different orders and cluster similar modes together. However, most proposed methods based on this approach suffer from high-dimensional optimization problems in either the estimation or clustering step. To overcome this problem, this study presents an algorithm for autonomous modal parameter estimation in which the only required optimization is performed in a three-dimensional space. To this end, a subspace-based identification method is employed for the estimation and a non-iterative correlation-based method is used for the clustering. This clustering is at the heart of the paper. The keys to success are correlation metrics that are able to treat the problems of spatial eigenvector aliasing and nonunique eigenvectors of coalescent modes simultaneously. The algorithm commences by the identification of an excessively high-order model from frequency response function test data. The high number of modes of this model provide bases for two subspaces: one for likely physical modes of the tested system and one for its complement dubbed the subspace of noise modes. By employing the bootstrap resampling technique, several subsets are generated from the same basic dataset and for each of them a model is identified to form a set of models. Then, by correlation analysis with the two aforementioned subspaces, highly correlated modes of these models which appear repeatedly are clustered together and the noise modes are collected in a so-called *Trashbox* cluster. Stray noise modes attracted to the mode clusters are trimmed away in a second step by correlation analysis. The final step of the algorithm is a fuzzy c-means clustering procedure applied to a three-dimensional feature space to assign a degree of physicalness to each cluster. The proposed algorithm is applied to two case studies: one with synthetic data and one with real test data obtained from a hammer impact test. The results indicate that the algorithm successfully clusters similar modes and gives a reasonable quantification of the extent to which each cluster is physical.

---

[1] To whom correspondence should be addressed. E-mail: *khorsand@chalmers.se*.

# 1  Introduction

## 1.1  Pertinent literature

Over the last decades, much effort has been put to develop efficient algorithms for identification of the modal parameters using time or frequency domain data [1, 2]. A central problem in most of these algorithms is to determine the true model order to capture the physical modes of the test-piece. However, this model order determination often demands considerable interaction from an experienced user. This hinders the use of developed modal analysis techniques for the applications which require a periodic estimation of the modal parameters like continuous health monitoring of structures.

In the framework of system identification, there exists an extensive literature for order estimation of linear dynamical models. The Akaike Information Criterion (AIC) [3] for maximum likelihood estimator and the Singular Value Criterion (SVC) [4] for subspace-based methods are two such examples of model order selection criteria. They share the idea of comparing the significance of the inclusion of yet another mode for increasing the prediction capability of the model with a penalty cost of including it. Such cost is somewhat sensitive to the choice of specific user parameters. Although these criteria can perform well for model validation in general, they often provide a slight overestimation of the model order [5, 6] and are also inadequate to detect and reject the physically irrelevant modes which often appear in the identified models [7]. Such irrelevant modes are here called noise modes without considering of their origin.

In the contrast, in the modal analysis community, the primary interest is often in the physical relevance of the individual modes of the identified model rather than a related model's prediction capacity. Therefore, the common practice is to identify a model with an order that is much higher than motivated by physics to ensure that all physical eigenmodes within the frequency band of interest are safely captured [8-10]. However, this inevitably results in the appearance of noise modes in the identified model, *i.e.,* modes which are present in the model due to measurement noise or computational imprecision but have no relevance to the physics of the tested system. Various tools have been developed to detect and eliminate such noise modes from a model. The most widespread tool is undoubtedly the so-called stabilization diagram [11, 12]. This diagram is constructed using estimated eigenfrequencies of models with increasing order. Ideally, for a physical mode, the estimated eigenfrequencies show up with the same value for increasing model order while for a noise mode they do not [6]. However, the interpretation of the stabilization diagram is an art which often requires a lot of user interaction. Specifically, for highly noisy data its outcome highly depends on user decisions.

In recent years, many studies attempted to automate the interpretation of stabilization diagram or the modal parameter estimation algorithm in general [13-17]. Owing to the fact that analyzing the stabilization diagram reduces to finding modes with similar properties, the majority of automation strategies borrow methods from statistical machine learning with supervised and unsupervised learning algorithms. Goethal *et al.* [12] proposed to utilize a supervised learning algorithm to automate the interpretation of stabilization diagrams. In their study, a hierarchical clustering algorithm groups similar modes of a stabilization diagram together. Then, the final decision on the nature of a cluster, being either physical or a noise artifact, is made by a self-learning Support Vector Machine (SVM) algorithm. Their hypothesis is that once the SVM algorithm is sufficiently trained from sets of data obtained from designed synthetic experiments, the algorithm will automatically classify physical modal parameters for real test data.

Special attention has been given to unsupervised learning algorithms. Hierarchical and centroid-based clustering[1] algorithms are two examples of this type of learning algorithms. Hierarchical clustering starts by assigning one cluster to each data point in a stabilization diagram. Then, it proceeds by merging the closest clusters together until the distance between the resulting clusters exceeds a user-defined threshold. Finally, physical modes are defined by clusters in which the number of modes is larger than a user-specified threshold. A considerable research effort has been made to develop appropriate distance measure for the hierarchical clustering. Magalhães *et al.* [18] suggested a distance measure which is based on the eigenfrequency difference and the Modal Assurance Criterion (MAC) value. Allemang *et al.* [19] used the MAC value between pole-weighted mode shapes as the distance measure between clusters. Goethals *et al.* [12] proposed a distance measure based on the difference of damping ratios and eigenfrequencies.

Other examples of unsupervised learning algorithms are centroid-based clustering schemes such as $k$-means and fuzzy $c$-means. In these clustering approaches, a central point/vector (centroid) serves as a representative for each cluster, although it may not be a member of data points in a stabilization diagram. Then the data points are grouped into $k$ clusters such that the squared distances from the cluster centroids are minimized. The main drawback of this approach is that the number of clusters is assumed to be known *a priori*, which is often not the case in practice. Scionti and Lanslots [20] employed fuzzy c-means clustering to directly group the existing modes in the stabilization diagram into a predefined number of clusters. Vanlanduit *et al.* [6] and Verboven *et al.* [8] proposed a frequency-domain Maximum Likelihood Estimator (MLE) to estimate the modal parameters using a single high model order $n$. Subsequently, they grouped the estimated modes into two classes of physical and noise modes using a fuzzy c-means clustering algorithm. Reynders *et al.* [7] automates the analysis of stabilization diagrams using three steps of clustering. First, a centroid-based clustering algorithm is employed to remove the noise modes from the stabilization diagram. Secondly, a hierarchical clustering algorithm is employed to group similar modes that survived the previous stage. Finally, they make use of a k-means clustering algorithm to group the resulting clusters in the last step into noise and physical clusters of modes.

---

[1] - In general, clustering refers to the task of subdividing a set of data points into subsets such that the (in some sense) similar data points are grouped together.

**1.2   Problem statement**

The ultimate goal of the present study is to develop a fully automated modal parameter estimation algorithm such that the following criteria are satisfied: (*i*) it should involve a system identification algorithm which allows for fast and robust identification of MIMO systems of a given order, (*ii*) it should avoid high-dimensional optimization, (*iii*) it should provide uncertainty bounds on the estimated modal parameters and (*iv*) it should need no user-specified parameters or thresholds.

In these respects, all modal parameter estimation methods discussed in the previous section have certain limitations that make them violate one or more of the aforementioned criteria. For instance, a vast majority of those methods require user-defined parameters or thresholds. To the authors' knowledge, the only exceptions are the algorithms proposed by Vanlanduit *et al.* [6], Reynder *et al.* [7], Verboven *et al.* [8], and Rainieri *et al.* [15, 17]. However, their methods involve high-dimensional optimization procedures, either for the estimation step [6, 8] or the clustering step [7], which deteriorates their performance both in terms of the computational efficiency and convergence. The method proposed in [15, 17] is suitable for modal tracking since it requires at least two datasets recorded under similar excitations.

Figure 1 shows the building blocks of the approach proposed to automate the estimation of modal parameters in input-output modal analysis. The five main steps constituting the proposed algorithm are as follows:

(I)   Identify a state-space model, $\mathbf{\Sigma}_e$, of exhaustively high order using a frequency-domain subspace-based identification method from noise corrupted measurements of the Frequency Response Function (FRF) $\boldsymbol{G}^*$. A subspace-based method is selected since it does not require expensive optimization of non-convex cost functions and thus does not suffer from associated convergence problems. We refer interested readers to Mckelvey *et al.* [21] for detailed proofs and discussions of the numerical stability of the method. From this a set of all modes that might be present in the frequency range of interest is established which is here called the *subspace of likely physical modes*. Its complementary subspace defines another set which provides the bases for another subspace which is called the *subspace of noise modes*.

(II) Identifying a set of state-space models, $\mathbf{\Sigma}_b$, $b = 1, \dots, N_\mathfrak{B}$, of order $n_r$ using repeated draws of $N_\mathfrak{B}$ bootstrap datasets from the measured FRFs. A conservatively high model order $n_r$ is determined by the SVC information criterion.

(III) Grouping similar modes of the repeated realizations using a centroid-based clustering with a correlation metric as the distance measure. The centroids of the clusters are the modes of the subspace of likely physical modes. The distances between realized modes and the centroids are given by a mode consistency metric that indicates the linear dependency between the modes and the centroids. The modes best correlated with the subspace of noise modes are clustered in a so-called *Trashbox*.

(IV) Trimming the likely physical clusters to remove stray noise modes which were caught by clusters in the preceding step. These noise modes are distinguishable from other modes of the cluster by their high correlation to the orthogonal complement of the dominant direction of the modes in the cluster.

(V) Fuzzy clustering to divide the clusters into two distinct classes, the physical modes class and the noise modes class. Three features are evaluated for the clusters and the classification is performed in this feature space to determine the degree to which each cluster is physical, *i.e.* Degree of Physicalness (DoP).

The development and validation of a generally applicable automated modal parameter estimation algorithm that satisfies all four aforementioned criteria are given in this paper. This paper also introduces new correlation metrics that are keys to the performance of the correlation-based clustering algorithm. It also describes the use of QR-decomposition to deal with coalescent eigenvalues.

It should be mentioned that this algorithm is designed for frequency domain data. Since data provided in the time domain can be transformed to the frequency domain, this imposes no restriction.

The remainder of this paper is organized as follows. Section 2 presents the two new correlation metrics that diminish the problem of existing correlation metrics to deal with spatial aliasing. Also, the problem of nonunique eigenvectors of coalescent modes is addressed here. It is demonstrated how a singular value decomposition (SVD) can be employed to automate the process of sorting out modes with less contribution from a group of modes. Section 3 explains different building blocks of the proposed automated modal parameter estimation algorithm in detail. Section 4 considers a synthetic dataset from an academic example for a validation study and also a real experimental dataset from a pentagon shape structure to illustrate the effectiveness of the proposed algorithm. Finally, Section 5 concludes the paper.

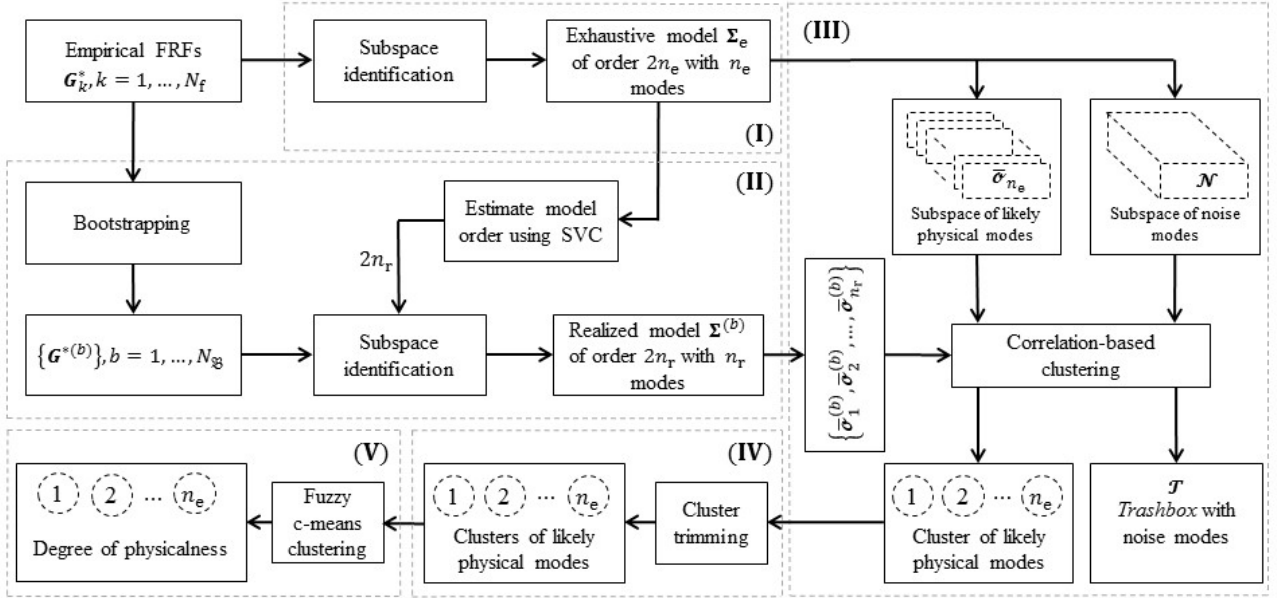

**Figure 1**. Flowchart of the proposed autonomous modal parameter estimation algorithm. Roman numbers refer to steps given in Section 1.2.

## 2 Correlation analysis

In this section, a correlation metric utilized for the correlation-based clustering procedure is described first. The pertinent properties of an SVD to rank the modes within a group of clustered modes based on their correlation to the dominant direction of the cluster is described.

### 2.1 Correlation metric: Modal Observability Correlation

Historically, the Modal Assurance Criterion (MAC) was developed as a metric of consistency between different estimates of a modal vector [22]. For systems with well-separated resonance frequencies and with many response locations for the representation of modal vectors, the MAC typically helps to separate modal vectors associated with different resonance frequencies. However, when few sensors are used for the experimental determination of modal vectors, high MAC correlations may occur between modal vectors of modes associated with different resonance frequencies. This is the spatial aliasing phenomenon. Furthermore, high MAC correlation can be due to the similarity between an eigenvector of a noise mode caused by measurement imperfection and a true system mode. The aforementioned problems motivated the development of the Modal Observability Correlation (MOC) metric, first proposed in [23].

The spatial aliasing effect arises when two modes of the structure are observed similarly through the system outputs. The relationship between the system's observability and modal properties is thus of special interest. To this end, let the continuous, linear and time-invariant (LTI) system $\Sigma = (A, B, C, D)$ be represented in a state-space form as

$$\begin{cases} \dot{x}(t) = Ax(t) + Bu(t) \\ y(t) = Cx(t) + Du(t) \end{cases} \quad (1)$$

where $x(t) \in \mathbb{R}^{n_x}$ is the state vector, $A \in \mathbb{R}^{n_x \times n_x}$, $B \in \mathbb{R}^{n_x \times n_u}$, $C \in \mathbb{R}^{n_y \times n_x}$ and $D \in \mathbb{R}^{n_y \times n_u}$ are internal evolution, input, output and feed-through matrices, respectively. The system input is $u(t) \in \mathbb{R}^{n_u}$ and the system output is $y(t) \in \mathbb{R}^{n_y}$. Then, in order to assess the observability of the system $\Sigma$ one can evaluate the rank of its associated observability matrix, $\mathcal{O}$, *i.e.*, the rank of

$$\mathcal{O} = \begin{bmatrix} C \\ CA \\ CA^2 \\ \vdots \\ CA^{n_x - 1} \end{bmatrix} \in \mathbb{R}^{(n_x \times n_y) \times n_x} \quad (2)$$

If $\mathcal{O}$ has full column rank, then the system is fully observable by the outputs $y$. If $A$ is non-deficient with the eigenvalue set $\lambda = \{\lambda_1, \lambda_2, \dots, \lambda_{n_x}\}$ and the corresponding eigenvector matrix $\Phi = [\phi_1, \phi_2, \dots, \phi_{n_x}]$, the transformation $\bar{x}(t) = \Phi^{-1} x(t)$ can be used to project the dynamical system $\Sigma$ onto a modal coordinate form as the quadruple $\bar{\Sigma} = (\bar{A}, \bar{B}, \bar{C}, D)$ where $\bar{A} = \Lambda = \text{diag}(\lambda_1, \lambda_2, \dots, \lambda_{n_x}) = \Phi^{-1} A \Phi$, $\bar{B} = \Phi^{-1} B = [\bar{b}_1, \bar{b}_2, \dots, \bar{b}_{n_x}]$ is the modally projected input matrix, and $\bar{C} = C\Phi = [\bar{c}_1, \bar{c}_2, \dots, \bar{c}_{n_x}]$ is the modally projected output matrix. The observability matrix related to this form can be written using the modal representation of the system as

$$\bar{\mathcal{O}} = \begin{bmatrix} C\Phi \\ C\Phi\Lambda \\ C\Phi\Lambda^2 \\ \vdots \\ C\Phi\Lambda^{n_x - 1} \end{bmatrix} = \begin{bmatrix} C\phi_1 & \cdots & C\phi_{n_x} \\ \vdots & \ddots & \vdots \\ C\phi_1 \lambda_1^{n_x - 1} & \cdots & C\phi_{n_x} \lambda_{n_x}^{n_x - 1} \end{bmatrix} \equiv [\bar{\sigma}_1 \; \cdots \; \bar{\sigma}_{n_x}] \quad (3)$$

This is the modal observability matrix and its columns are the modal observability vectors. As this shows, each modal observability vector, $\bar{\sigma}_i, i = 1, 2, \dots, n_x$, has the following important properties: (*i*) it contains information related to only one single vibrational mode, *i.e.*, it is a modal characteristic vector, (*ii*) it includes both eigenvector and eigenvalue information, (*iii*) it relates the modal parameters to the observability of the modes given by a used sensor configuration. These properties suggests a departure from the traditional MAC formulation which only uses eigenvector information and replace its eigenvectors by the modal observability vectors. In analogy with the definition of MAC [22], for the correlation between two modal observability vectors $\bar{\boldsymbol{\sigma}}_i$ and $\bar{\boldsymbol{\sigma}}_j$, we thus have

$$b\text{MOC}(\bar{\boldsymbol{\sigma}}_i, \bar{\boldsymbol{\sigma}}_j) = \frac{|\bar{\boldsymbol{\sigma}}_i^\text{H} \bar{\boldsymbol{\sigma}}_j|^2}{(\bar{\boldsymbol{\sigma}}_i^\text{H} \bar{\boldsymbol{\sigma}}_i)(\bar{\boldsymbol{\sigma}}_j^\text{H} \bar{\boldsymbol{\sigma}}_j)}, \quad i,j = 1,2,\ldots,n_x \qquad (4)$$

We call this metric the basic Modal Observability Correlation ($b$MOC). Since the vectors $\bar{\boldsymbol{\sigma}}_i$ and $\bar{\boldsymbol{\sigma}}_j$ contain information about the eigenvectors and the associated eigenvalues, this metric allows for discriminating between two highly correlated modal vectors with non-coinciding eigenvalues. This makes $b$MOC better than MAC in dealing with the problem of spatial aliasing.

To avoid numerical problems with raising $\Lambda$ to high exponents, see Eq. (3), the $b$MOC is evaluated for discrete-time system eigenvalues. From continuous-time systems, the eigenvalues $\lambda_j, j = 1, 2, \ldots, n_x$, are transformed to the discrete-time counterpart using $\lambda_{j,\text{disc}} = \exp(\lambda_j \tau)$ where $\tau$ is an arbitrary time step. One choice is to let $\tau = \pi/f_{\max}$ in which $f_{\max}$ is the maximum frequency used for the FRF. This leads to a distribution of the discrete-time eigenvalues $\lambda_{j,\text{disc}}$ over a large domain of the complex plane unit disc and thus separates them well.

Since the $b$MOC factors are based on the columns of the modal observability matrix normalized by their magnitude, it only measures vector collinearity and the vector norms plays no role. This means that large $b$MOC correlation could be obtained between two modal vectors which contribute very differently to the system's input-output relation. A true system mode of significant relevance can then easily be taken as correlated to a mode which only represent low level measurement noise. To address this issue, one should notice that the noise modes often give small contributions to the input-output relation of the model [11]. In other words, they can be revealed by their small controllability and/or observability. This advocates the use of a transformation in which the modal scaling reflects the mode's degree of controllability and observability. To this end, let the decoupled subsystem associated to each eigenvalue $\lambda_j$ be written as

$$\begin{cases} \dot{\bar{x}}_j(t) = \lambda_j \bar{x}_j(t) + \bar{\boldsymbol{b}}_j \boldsymbol{u}(t) \\ \boldsymbol{y}_j(t) = \bar{\boldsymbol{c}}_j \bar{x}_j(t) + \boldsymbol{D}\boldsymbol{u}(t) \end{cases} \qquad (5)$$

Using the balancing transformation $\tilde{T} \in \mathbb{R}^1$, the quadruple $(\lambda_j, \tilde{\bar{\boldsymbol{b}}}_j, \tilde{\bar{\boldsymbol{c}}}_j, \boldsymbol{D}) = (\lambda_j, \tilde{T}^{-1}\bar{\boldsymbol{b}}_j, \bar{\boldsymbol{c}}_j\tilde{T}, \boldsymbol{D})$ is then the internally balanced representation of the system associated to the $j$th mode. The balancing makes it equally controllable and observable, *i.e.*,

$$\|\tilde{T}^{-1}\overline{\boldsymbol{b}}_j\| = \|\overline{\boldsymbol{c}}_j\tilde{T}\| \tag{6}$$

Here, $\|\cdot\|$ denotes the norm of a vector. Then, the $j$th modal observability vector can be written as

$$\widetilde{\overline{\boldsymbol{\sigma}}}_j = \tilde{T}\overline{\boldsymbol{\sigma}}_j. \tag{7}$$

Note that, even though this transformation will not change the $b$MOC number, *i.e.*, for $i$th and $j$th modes $b\text{MOC}(\overline{\boldsymbol{\sigma}}_i, \overline{\boldsymbol{\sigma}}_j) = b\text{MOC}(\widetilde{\overline{\boldsymbol{\sigma}}}_i, \widetilde{\overline{\boldsymbol{\sigma}}}_j)$, its uniqueness property renders a meaningful and unique scaling of the modal vectors such that their norms now indicate the extent to which they contribute to the system's input-output behavior. This is a valuable information which will help us later in Section 3.4. to rank the importance of the modes in one set of correlated modes. Therefore, throughout this paper, the modal observability vector implies its balanced transform and thus $\widetilde{\overline{\boldsymbol{\sigma}}}_j$ replaces with $\overline{\boldsymbol{\sigma}}_j$ for simplicity.

In fact, the idea of using the information of eigenvalues in consistency metrics can be traced back to the study by Phillips and Allemang [24] which introduced the pole-weighted MAC (pwMAC). This consistency metric is namely the MAC value between the so-called pole-weighted modal vectors, which can be viewed as the modal observability vector truncated at a certain power of the eigenvalues. In the following, the concept of auto-MOC is used in analogy with the well-known auto-MAC. This means that the vector set used for correlation analysis stems from a single source that could be either from an experimental modal analysis or from a finite element based eigensolution.

### 2.1.1 Dealing with coalescent eigenvalues

One of the major challenges in modal analysis is that of coalescent or closely spaced eigenvalues. Such eigenvalues may relate to noise modes or physical system modes. In the case of physical modes, identified multiple eigenvalues can be an artifact of a too high model orders tried in an estimation step or they may correspond to truly multiple modes of the system. Thus, clustering-based approaches need appropriate tools to detect and separate multiple modes into distinct clusters before a correlation analysis takes place. The QR-decomposition is used here to establish a fixed orthogonal basis in the invariant subspace of multiple modes of an identified model. Let $\boldsymbol{\Phi}_j \in \mathbb{C}^{n_y \times n_c}$ represent a matrix which contains the linear independent eigenvectors associated to a coalescent eigenvalue $\lambda_j$ of multiplicity $n_c$. Theoretically, the dimension $n_c$ of multiple eigenvalues of an identified model cannot exceed the number of actuators, $n_u$ for controllability reasons. The $n_c$-dimensional space spanned by the columns of $\boldsymbol{\Phi}_j$, $\boldsymbol{\mathcal{V}}$, is an $\boldsymbol{A}$-invariant[2] subspace of $\mathbb{C}^{n_y}$ where any vector that lies in this subspace can be considered to be an eigenvector corresponding to $\lambda_j$. Different variants of a linearly independent basis for $\boldsymbol{\mathcal{V}}$ may then appear in repeated identification made from subsets of FRF drawn from a given dataset. To treat this issue, the interest is thus to search for a fixed orthogonal basis for $\boldsymbol{\mathcal{V}}$. The unique projection of a vector using the system input matrix $\boldsymbol{B}$ is used to establish a fixed orthogonal basis for $\boldsymbol{\mathcal{V}}$. The projection of the columns of $\boldsymbol{B}$ onto $\boldsymbol{\mathcal{V}}$, $\boldsymbol{P}_{\boldsymbol{\mathcal{V}}}(\boldsymbol{B})$, can be defined as

$$\boldsymbol{P}_{\boldsymbol{\mathcal{V}}}(\boldsymbol{B}) = \boldsymbol{\Phi}_j (\boldsymbol{\Phi}_j^{\mathrm{H}} \boldsymbol{\Phi}_j)^{-1} \boldsymbol{\Phi}_j^{\mathrm{H}} \boldsymbol{B} \tag{8}$$

where $\boldsymbol{P}_{\boldsymbol{\mathcal{V}}}(\boldsymbol{B}) \in \mathbb{C}^{n_x \times n_\mathrm{u}}$. The QR-decomposition of $\boldsymbol{P}_{\boldsymbol{\mathcal{V}}}(\boldsymbol{B})$ gives

$$\boldsymbol{P}_{\boldsymbol{\mathcal{V}}}(\boldsymbol{B}) = \boldsymbol{QR} \tag{9}$$

Here, $\boldsymbol{Q} \in \mathbb{C}^{n_x \times n_x}$ is a unitary matrix and $\boldsymbol{R} \in \mathbb{C}^{n_x \times n_\mathrm{u}}$ is an upper triangular matrix. The first $n_\mathrm{c}$ columns of $\boldsymbol{Q}$ forms the desired fixed orthogonal basis for $\boldsymbol{\mathcal{V}}$, *i.e.*,

$$\boldsymbol{\mathcal{V}} = \mathrm{span}\{\boldsymbol{Q}_1, \boldsymbol{Q}_2, \cdots, \boldsymbol{Q}_{n_\mathrm{c}}\} \tag{10}$$

The space spanned by the columns $\boldsymbol{Q}_i$ is also $\boldsymbol{A}$-invariant and thus the unique vectors $\boldsymbol{Q}_i, i = 1,2,\dots,n_c$ are eigenvectors associated to the coalescent eigenvalues, $\lambda_j$ [25]. The problem of eigenvector nonuniqueness for coalescent eigenvalues is thus solved by the QR-decomposition. This decomposition will be employed throughout this study to fix the eigenvectors of coalescent modes which may appear in the identification stage.

## 2.2   Complementary subspace correlation analysis

Different modal parameter estimation algorithms working on the same dataset does not provide exactly the same eigensolutions. Also, one single algorithm working on different subsets of data from the same dataset provide different modal parameter estimates. Therefore, in high model order estimates, there is a presence of some eigenvectors reflecting the random noise in the data or numerical noise or bias given by the execution of the modal parameter estimation algorithm. These modes often show a negligible degree of consistency with the other estimated modes. In a correlation-based clustering algorithm, it is desirable to form a set of basis vectors which exhibits a noticeable degree of consistency with such randomly appearing modal vectors.

---

[2] If $\boldsymbol{F} \in \mathbb{C}^{p \times q}$ is a matrix whose columns $\boldsymbol{f}_1, \boldsymbol{f}_2, \dots, \boldsymbol{f}_q$ span a $q$-dimensional subspace, $\boldsymbol{\mathcal{F}}$. Then, $\boldsymbol{\mathcal{F}}$ is $\boldsymbol{A}$-invariant if and only if there exists a matrix $\boldsymbol{Z} \in \mathbb{C}^{q \times q}$ that satisfies $\boldsymbol{AF} = \boldsymbol{FZ}$.

To this end, let $\overline{\boldsymbol{O}}_{n_s} = [\overline{\boldsymbol{\sigma}}_1, \cdots, \overline{\boldsymbol{\sigma}}_{n_s}] \in \mathbb{C}^{(n_x \times n_y) \times n_s}$ denote a matrix containing $n_s$ modal observability columns $\overline{\boldsymbol{\sigma}}_i$ associated to $n_s$ estimated modal vectors. These modal vectors can represent the set of $n_s$ modal vectors estimated for a system, as described later in Section 3.1, or represent different estimates of the same modal vector, see Section 3.4. In both cases, our interest is to find the bases for the orthogonal complement of the space spanned by the dominant basis vectors of $\overline{\boldsymbol{O}}_{n_s}$. Considering its singular value decomposition, $\overline{\boldsymbol{O}}_{n_s} = \boldsymbol{USV}^{\mathrm{H}}$, the matrices $\boldsymbol{U} \in \mathbb{C}^{(n_x \times n_y) \times (n_x \times n_y)}$, $\boldsymbol{S} \in \mathbb{R}^{(n_x \times n_y) \times n_s}$ and $\boldsymbol{V} \in \mathbb{C}^{n_s \times n_s}$ can be partitioned into two sets as

$$\overline{\boldsymbol{O}}_{n_s} = \boldsymbol{USV}^{\mathrm{H}} = [\boldsymbol{U}_1 \quad \boldsymbol{U}_2] \begin{bmatrix} \boldsymbol{S}_1 & 0 \\ 0 & \boldsymbol{S}_2 \end{bmatrix} \begin{bmatrix} \boldsymbol{V}_1^{\mathrm{H}} \\ \boldsymbol{V}_2^{\mathrm{H}} \end{bmatrix} \tag{11}$$

where $\boldsymbol{S}_1 = \mathrm{diag}(\sigma_1, \dots, \sigma_r) \in \mathbb{R}^{r \times r}$ of the first set contains the $r$ largest non-zero singular values of the matrix $\overline{\boldsymbol{O}}_{n_s}$ and the size of submatrices are all determined by $r$, e.g., $\boldsymbol{U}_1 \in \mathbb{C}^{(n_x \times n_y) \times r}$ and $\boldsymbol{V}_1 \in \mathbb{C}^{n_s \times r}$. Given Eq. (11), the following properties hold:

(1) If $\overline{\boldsymbol{O}}_{n_s}$ represents the set of $n_s < n_x \times n_y$ modal vectors estimated for a system, and $r$ is such that $\sigma_1 \geq \sigma_2 \geq \cdots \geq \sigma_r > \sigma_{r+1} = \cdots = \sigma_{n_s} = 0$. Then, the columns of $\boldsymbol{U}_1$ provide an $r$-dimensional orthogonal basis for the dominant range space of $\overline{\boldsymbol{O}}_{n_s}$ and the columns of $\boldsymbol{U}_2$ form its orthogonal complement subspace, $\boldsymbol{\mathcal{U}}$.

(2) If $\overline{\boldsymbol{O}}_{n_s}$ represents different estimates of the same modal vector, we let $r = 1$ and thus $\boldsymbol{S}_1 = \sigma_1$ holds only the largest singular value of $\overline{\boldsymbol{O}}_{n_s}$. Therefore, $\boldsymbol{U}_1$ contains the dominant basis vector for the modal observability columns in the matrix $\overline{\boldsymbol{O}}_{n_s}$, and the columns of $\boldsymbol{U}_2$ span its orthogonal complement subspace, $\boldsymbol{\mathcal{U}}$.

Let $\overline{\boldsymbol{\sigma}}_k$ denotes a modal observability vector that shows low correlation with the basis of the dominant subspace of $\overline{\boldsymbol{O}}_{n_s}$, then, one can expect a high correlation between $\overline{\boldsymbol{\sigma}}_k$ and the orthogonal complement subspace of $\overline{\boldsymbol{O}}_{n_s}$, i.e., $\boldsymbol{\mathcal{U}}$. To perform such correlation analysis between a vector and a subspace, we introduce another correlation metric called *higher order MOC* ($h$MOC). This metric can be seen as an extension of $b$MOC and is

$$h\mathrm{MOC}(\boldsymbol{\mathcal{U}}, \overline{\boldsymbol{\sigma}}_k) = \cos^2(\sphericalangle[\boldsymbol{\mathcal{U}}, \overline{\boldsymbol{\sigma}}_k]) \tag{12}$$

Here, $\sphericalangle$ denotes the angle between the vector $\overline{\boldsymbol{\sigma}}_k$ and the subspace $\boldsymbol{\mathcal{U}}$. A $h$MOC value indicates the extent to which the vector $\overline{\boldsymbol{\sigma}}_k$ belongs to the subspace $\boldsymbol{\mathcal{U}}$. This correlation metric ranges from 0 to 1. A correlation number at 0 means that the vector $\overline{\boldsymbol{\sigma}}_k$ is orthogonal to the subspace. On the other hand, if the correlation number is close to 1 it can be well described by the subspace $\boldsymbol{\mathcal{U}}$. We then loosely say that it belongs to the subspace $\mathcal{U}$. Later, the concepts of the complementary subspace and the associated $h$MOC will be used in the correlation-based clustering algorithm (Section 3.1), and also used to remove noise modes from clusters (Section 3.4).

## 3 Autonomous modal parameter estimation

Assume that noise corrupted samples of the frequency response function, $\{G_k^*\}, k = 1, \ldots, N_f$ are available. Here $G_k^* \in \mathbb{C}^{n_y \times n_u}$ and $N_f$ is the number of discrete frequencies for which $G^*$ is known. The steps of the proposed algorithm for estimation of the modal parameters from $\{G_k^*\}$ are explained thoroughly in this section. The steps are shown as blocks in the flowchart presented in Figure 1.

### 3.1 Exhaustive model

As shown in block (I) of Figure 1, the algorithm commences with the estimation of a linear state-space model $\Sigma_e$ with a very high order $2n_e$ using the subspace-based identification algorithm [21]. This model is called the exhaustive model (EM) hereafter. The order $2n_e$ is selected to be much higher than necessary to capture all pertinent physical characteristics of the test structure in the considered frequency range. Only the $n_e$ modes with positive imaginary eigenvalues are considered. From the model $\Sigma_e$ a set of these $n_e$ modes are obtained and characterized by their natural frequency $\omega_i$, modal damping $\xi_i$, mode shape $\boldsymbol{\varphi}_i$ and modal observability vector $\bar{\boldsymbol{\sigma}}_i$, $i = 1, \ldots, n_e$. Two significant roles of the exhaustive model $\Sigma_e$ are to (*i*) form bases for the subspace of likely physical modes and (*ii*) form bases for the subspace of noise modes. This will be elaborated on in the following.

#### 3.1.1 Subspace of likely physical modes

Since the EM is identified with an extensively high model order it contains noise modes as well as physical modes. Therefore, the subspace spanned by the columns of the modal observability matrix $\overline{\boldsymbol{\mathcal{O}}} \in \mathbb{R}^{(n_e \times n_y) \times n_e}$ is the union of the two subspaces that span the physical modes and the noise modes. These subspaces are not easily distinguishable beforehand and therefore, we call their union the subspace of likely physical modes. Each mode of the EM can be a representative (or centroid) for one cluster of likely physical modes. Later, we will show that one can scan over these centroids in order to group modes obtained from bootstrap realizations with the centroid to which they show the highest correlation.

#### 3.1.2 Subspace of noise modes

Since the EM has a high model order, the modal observability matrix associated with this model, $\overline{\boldsymbol{\mathcal{O}}}$, very likely contains all physical modes of the test structure which are observable from the sensor configuration. Therefore, see Section 2.2, it is expected that its complementary subspace $\mathcal{N}$, i.e., $\mathcal{N} = \mathcal{U} \in \mathbb{R}^{(n_e \times n_y) \times (n_e \times (n_y-1))}$ is the subspace spanned by the columns of $U_2$, gives high correlation with any mode that show low correlation with the centroids of the likely physical clusters. Such modes for which no physical relevance related to the test-piece is expected are called noise modes and this subspace is called the subspace of noise modes. The cluster associated to this subspace is called the *Trashbox* $\mathcal{T}$. Correlation analyses between modal observability vectors $\bar{\sigma}_k$ with the subspace $\mathcal{N}$, is made using the higher order $b$MOC ($h$MOC).

## 3.2 Bootstrapping

The subspace-based identification algorithm used here does not provide uncertainty bounds on the estimated modal parameters. Such bounds could give valuable and decisive information for discriminating between physical and noise modes. The second step of the algorithm, shown as the block (II) in Figure 1, is devised to provide such information by the use of the bootstrapping method. Bootstrapping is one of the statistical techniques operating on the measured data to infer a level of uncertainty on any parametric estimator [26]. The idea behind the bootstrapping is to assess the statistical property of an estimator, such as its mean or variance, by calculating it from repeated random sampling of the same given dataset. It can be implemented by randomly drawing samples with replacement of the observed dataset $N_\mathcal{B}$ times. This leads to $N_\mathcal{B}$ datasets of the same size as the original dataset but with different subsets of data taken every time. For each such sampling a state-space model is estimated and then, the statistical behavior of these models can be examined (see [27] for further information). Evaluating statistics for the estimated modal parameters using the bootstrapping method can be exceedingly expensive if the exhaustive model order $2n_e$ would apply to each bootstrap dataset. This motivates the need for a model order smaller than $2n_e$ but still higher than the number of physical modes of the system. To this end, the $SVC$ is adopted here.

### 3.2.1 Singular value criterion

The Singular Value Criterion ($SVC$) is an information theoretic criterion proposed in [4] to estimate the model order, $2n_r$, in the context of subspace-based identification. It is adopted here to find a proper order of models fitted to the bootstrap datasets. This criterion metric is defined as

$$SVC(n) = \sigma_{n+1}^2 + \frac{\mathcal{C}(N_d)d(n)}{N_d}, \qquad n = 1, 2, \dots, 2n_e \qquad (13)$$

where $d(n) = n(n_u + 2n_y) + n_y n_u$ is the minimum number of parameters of a state-space model of order $n$, $\sigma_{n+1}$ is the estimated Hankel singular values of the system's balanced Gramian, $N_d$ is the number of given data points and $C(N_d)$ is a model complexity penalty factor which will be described below. A low $SVC$ value indicates a good balance between model fit to data and model order since by increasing the model order $n$ the first term of the metric decreases and the second term increases. The model order which minimizes this criterion function is thus selected as an estimate of the proper model order $2n_r$. The model complexity factor is $C(N_d) = k_{c1}k_{c2} \log N_d$ for which $k_{c1}$ and $k_{c2}$ are integer numbers. From theoretical considerations and a practical point of view, as discussed in [4], appropriate choices of $k_{c1}$ and $k_{c2}$ are $k_{c1} = k_{c2} = d\hat{p}_{\text{AIC}}$ where $d > 1$ and $\hat{p}_{\text{AIC}}$ is the estimated model order by the Akaike Information Criterion (AIC) [3]. In this paper, the same idea is adapted with $d = 1$ to give a conservative overestimation of the model order and $\hat{p}_{\text{AIC}}$ is set to the number of complex conjugate mode pairs of the exhaustive model. Thus, $k_{c1} = k_{c2} = n_e$ is used here.

*3.2.2 Parameters statistics by bootstrapping*

In this study, a set of bootstrap datasets, $\{G^{*(b)}\}, b = 1, \dots, N_\mathcal{B}$ is collected from the noise corrupted frequency response function $G_k^*, k = 1, \dots, N_f$. Subsequently, a set of linear state-space models, $\Sigma_b, b = 1, \dots, N_\mathcal{B}$, of order $2n_r$ is estimated by applying the subspace-based identification method to each bootstrap dataset. In analogy with the EM, only the $n_r$ modes with positive imaginary part are kept for analysis. Then, the following properties can be extracted for each of these modes.

$$\begin{aligned} \omega_j^{(b)} &= \Im(\lambda_j^{(b)}) \\ \xi_j^{(b)} &= -\Re(\lambda_j^{(b)})/|\lambda_j^{(b)}| \\ \boldsymbol{\psi}_j^{(b)} &= \overline{\boldsymbol{C}}^{(b)}(:,j) = \overline{\boldsymbol{c}}_j^{(b)} \\ \overline{\boldsymbol{\sigma}}_j^{(b)} &= \overline{\boldsymbol{\mathcal{O}}}^{(b)}(:,j) \end{aligned} \qquad j = 1,2,\dots,n_r \qquad (14)$$

Here, $\omega_j^{(b)}, \xi_j^{(b)}, \boldsymbol{\psi}_j^{(b)}$ and $\overline{\boldsymbol{\sigma}}_j^{(b)}$ are the $j$th natural frequency, damping coefficient, mode shape and modal observability vector of $\Sigma_b$, respectively. $\Im(\circ)$, $\Re(\circ)$ and $|\circ|$ denote imaginary part, real part and magnitude of a complex number. Furthermore, another important property is the maximum modal contribution of the $j$th mode to the input-output relation which can be quantified as follows [28],

$$MC_j^{(b)} = \max_{\omega \in \mathbb{R}} \left| \frac{\overline{\boldsymbol{c}}_j^{(b)} \overline{\boldsymbol{b}}_j^{(b)}}{i\omega - \lambda_j^{(b)}} \right|, \qquad j = 1,2,\dots,n_r \qquad (15)$$

here $i^2 = -1$. Any aspect of the empirical distribution of these estimated modal parameters can then be estimated by bootstrapping. For instance, the variance of the estimated modal damping can be estimated by

$$\widehat{var}[\xi_j] = \frac{1}{N_\mathcal{B} - 1} \sum_{b=1}^{N_\mathcal{B}} \left(\xi_j^{(b)} - \hat{\mu}(\xi_j)\right)^2 \tag{16}$$

where $\hat{\mu}(\xi_j)$ denotes the estimated expected value of the $j$th damping coefficient which can be evaluated as

$$\hat{\mu}(\xi_j) = \frac{1}{N_\mathcal{B}} \sum_{b=1}^{N_\mathcal{B}} \xi_j^{(b)} \tag{17}$$

### 3.3 Correlation-based clustering

In general, the target of clustering is to put a set of objects into different groups (or clusters) such that the objects in a cluster are in some sense more similar to each other than to objects of other clusters. The objects are characterized by their features. However, the measure of similarity often cannot be uniquely defined by the object features which is one reason for the development of various clustering algorithms in the statistical data analysis and machine learning fields.

A straightforward procedure for clustering of vibrational modes is proposed here. As shown in block (III) of Figure 1, each cluster is represented by one cluster centroid. For likely physical modes, the $n_e$ modal observability vectors of the EM play role as the clusters' centroid whereas, the whole subspace $\mathcal{N}$ become a centroid for one cluster to collect the noise modes. Hence, the total number of clusters is set to, $n_e + 1$. In the clustering procedure, each mode of a Bootstrapped Model (BM) is assigned to that cluster whose centroid shows the highest level of correlation to this mode. The correlation metrics used in this step are the $b$MOC and the $h$MOC. Therefore, given a set of modes together with a set of modal observability vectors, $\bar{\sigma}_p^{(b)}, p = 1, \ldots, n_r$, corresponding to the $b$th BM, the proposed clustering algorithm group a mode into the cluster $c_i, i = 1, \ldots, n_e$, such that

$$b\text{MOC}\left(\bar{\sigma}_p^{(b)}, \bar{\sigma}_i\right) > \left\{b\text{MOC}\left(\bar{\sigma}_p^{(b)}, \bar{\sigma}_j\right), h\text{MOC}\left(\bar{\sigma}_p^{(b)}, \mathcal{N}\right)\right\}, \forall j: 1 \leq j \leq n_e, j \neq i \tag{18}$$

or group it to the *Trashbox* $\mathcal{T}$ if

$$b\text{MOC}\left(\bar{\sigma}_p^{(b)}, \bar{\sigma}_j\right) < h\text{MOC}\left(\bar{\sigma}_p^{(b)}, \mathcal{N}\right), \forall j: 1 \leq j \leq n_e. \tag{19}$$

This step is repeated for all BMs, $\mathbf{\Sigma}_b, b = 1, \ldots, N_\mathcal{B}$. The most apparent advantages of this algorithm are (*i*) this procedure can be implemented very efficiently due to the simplicity of the decision process in the assignment step, (*ii*) the algorithm is not iterative as is the case for instance for typical $k$-means algorithms and (*iii*) the proposed algorithm does not rely on any user-defined parameters or thresholds. At the end of this step, the clusters which did not attract any member and thus, are left empty are eliminated and the number of remaining likely physical clusters $n_e$ is adjusted accordingly.

## 3.4 Cluster trimming

Let the columns of $\overline{\mathbf{O}}_{c_i} = \begin{bmatrix} \overline{\boldsymbol{\sigma}}_1 & \cdots & \overline{\boldsymbol{\sigma}}_{N_{m_i}} \end{bmatrix}$ represent a set of modal observability vectors associated to the $N_{m_i}$ modes assigned to the cluster $c_i$, $i = 1,2, \ldots, n_e$. Further let the SVD of $\overline{\mathbf{O}}_{c_i}$ using Eq. (11) be used to obtain its singular values and its associated left singular vectors in $\mathbf{U}_1 \in \mathbb{C}^{(n_x \times n_y) \times r}$ and $\mathbf{U}_2 \in \mathbb{C}^{(n_x \times n_y) \times r}$. Note that the number of significant singular values $r$ represents the number of significant modal observability vectors in the cluster. Ideally, we expect each cluster to contain only repeated estimations of one single physical mode meaning that $\sigma_1 \gg \sigma_2 = \cdots = \sigma_{N_{m_i}} = 0$. However, two exceptions can happen which need to be treated differently.

Firstly, multiple modes of a coalescent eigenvalue of a BM can be clustered together [29]. This problem can be mitigated by utilizing the QR-decomposition to rotate the eigenvectors such that they become orthogonal to each other by a fixing projection using the input matrix $\mathbf{B}$, see Section 2.1.1.

Secondly, in spite of the fact that the *Trashbox* $\mathcal{T}$ collects noise modes of the BMs, a noise mode may give better correlation with a cluster centroid than with the subspace $\mathcal{N}$ and thus be assigned to its associated cluster. This in turn leads to the presence of noise modes in the clusters. These modes can often be distinguished by their relatively low correlation with the other modes in the cluster, see *e.g.*, Figure 8a. Such noise modes have to be trimmed away from the clusters in order to obtain valid statistics for the estimated modal parameters. In presence of such modes the sequence of singular values of $\overline{\mathbf{O}}_{c_i}$ change to $\sigma_1 \gg \sigma_2 \geq \cdots \geq \sigma_{N_{m_i}} \geq 0$ with $\mathbf{U}_1$ and $\mathcal{U}$ representing the dominant modal observability vector and its complementary orthogonal subspace, respectively (see Section 2.1.1). We propose to keep only the $N_{cm_i} \leq N_{m_i}$ modes that meet the following condition:

$$b\text{MOC}(\mathbf{U}_1, \overline{\boldsymbol{\sigma}}_k) > h\text{MOC}(\mathcal{U}, \overline{\boldsymbol{\sigma}}_k), \qquad k = 1, \ldots, N_{m_i} \qquad (20)$$

Modes that do not meet this criterion are trimmed away from the clusters to the *Trashbox* $\mathcal{T}$.

An additional constraint is imposed by the construction of the clusters. The upper bound on $N_{cm_i}$ is based on the fact that a cluster ideally cannot contain more modes than the number of bootstrap realizations $N_\mathfrak{B}$. A cluster with $N_\mathfrak{B}$ modes can thus be considered as being full. This means the clusters for which still $N_{cm_i} > N_\mathfrak{B}$ after the trimming operation (20) are overfull and therefore, some more modes should be trimmed away from each such cluster such that only the $N_\mathfrak{B}$ most relevant modes remain in the cluster. A measure for such modal relevancy is developed in the following.

As discussed in Section 2.1, the $b\text{MOC}(\boldsymbol{U}_1, \bar{\boldsymbol{\sigma}}_k)$ is a only measure of vector collinearity. This means, it only indicates how much $\bar{\boldsymbol{\sigma}}_k$ is parallel with $\boldsymbol{U}_1$ regardless of the magnitude of $\bar{\boldsymbol{\sigma}}_k$, although the magnitude is an important criteria for measuring the modal relevancy. In order to take it into account, it is proposed to use the projection of $\bar{\boldsymbol{\sigma}}_k$ onto the dominant basis instead, as

$$P_{\bar{\boldsymbol{\sigma}}_k} = \|\bar{\boldsymbol{\sigma}}_k\|^2 b\text{MOC}(\boldsymbol{U}_1, \bar{\boldsymbol{\sigma}}_k), \quad k = 1, \dots, N_{m_i} \tag{21}$$

This metric is defined as the modal relevancy and since $\bar{\boldsymbol{\sigma}}_k$ has been balanced, the vector norm is meaningful and unique and indicates the extent to which its associated mode contributes to the system input-output relation. With this metric, the clustered modes can be ranked based on their correlation to the dominant basis weighted by their contribution to the input-output relation. At the maximum, $N_\mathfrak{B}$ modes with the highest $P_{\bar{\boldsymbol{\sigma}}_k}$ will be kept in each cluster. The number of modes kept define the cluster size $N_{m_i}$.

## 3.5 Mode classification

In step V, the available clusters are classified into two classes: the physical modes class and the noise modes class. This step consists of three stages: (*i*) feature evaluation, (*ii*) interval normalization and (*iii*) iterative fuzzy c-means clustering. The mode classification should be based on discriminative features. This means, the features should be ideally such that they are very different for physical modes and noise modes. Given the number of remaining clusters $n_e$, the features space considered here consists of three features:

*The mean of the modal contributions, $\hat{\mu}(MC_i)$*

The fact that modes worth considering as physical should have relevance to the system dynamics, and thus contribute more to the input-output relation than the noise modes, makes the modal contribution an important feature to discriminate between physical and noise modes [30]. Therefore, the mean of the contribution of the modes grouped in one cluster can be used as a feature. However, since this feature has a wide range of variation, its logarithm is a better alternative. In order to discriminate between clusters with the same contribution mean but with different number of modes $N_m$, this feature is weighted by $N_m$. The contribution feature of the *i*th cluster is thus $\gamma_1^{(i)} = N_{m_i} \log(\hat{\mu}(\text{MC}_i))$.

*The coefficient of variation of eigenfrequency, $\widehat{CoV}(\omega_i)$*

It can be expected that the standard deviation of an eigenfrequency is small for a cluster of physical modes and larger for a cluster of noise modes. In repeated estimation of the modal parameters, the standard deviation of the eigenvalues of the noise modes is usually $10 - 100$ times larger than that of the physical modes, see [8]. Therefore, its associated coefficient of variation, $\widehat{CoV}(\omega_i) = \frac{\sqrt{\widehat{Var}(\omega_i)}}{\hat{\mu}(\omega_i)}$, can be used as another feature to discriminate between the physical and noise mode clusters. However, since small $\widehat{CoV}(\omega_i)$ is an indication to a high level of physical significance, it is inversely proportional to the Degree of Physicalness (DoP), *i.e.* the degree to which a cluster belongs to the physical mode class, its inverse is used here instead. The wide range of variations of this inverse motivates the use of logarithmic metric. Thus the eigenfrequency variation feature of the *i*th cluster is $\gamma_2^{(i)} = \log\left(\widehat{CoV}^{-1}(\omega_i)\right)$.

*The mean of modal damping, $\hat{\mu}(\xi_i)$*

In most practical modal testing application, modes with very high level of damping are rarely encountered and damping ratios larger than 20% can be considered as non-realistic [7]. Therefore, in such applications, the mean of modal damping of the clustered modes is a decisive criterion to distinguish between physical and noise modes. Since this feature is inversely proportional to the DoP, its inverse will be used as the feature $\boldsymbol{\gamma}_3$. Thus the damping feature of the $i^{th}$ cluster is $\gamma_3^{(i)} = \hat{\mu}^{-1}(\xi_i)$.

For classification, the features have to be normalized such that the features of the same type have the same range [31]. To this end, each positive real feature vector $\boldsymbol{\gamma}_l = \left[\gamma_l^{(1)}, \gamma_l^{(2)}, \dots, \gamma_l^{(n_e)}\right]^T$, $l = 1, 2, 3$ is subtracted by its minimum and the resulting difference is normalized with respect to its maximum. Therefore, all features are scaled to the interval $[0, 1]$ such that when they tend to unity, they represent a high level of physical relevance. On the other hand, if they tend to zero they represent some artificial modes that we consider to be un-physical or noise.

Let the feature space define as

$$\boldsymbol{\Gamma} = [\boldsymbol{\gamma}_1, \boldsymbol{\gamma}_2, \boldsymbol{\gamma}_3] = \begin{bmatrix} \gamma_1^{(1)} & \gamma_2^{(1)} & \gamma_3^{(1)} \\ \gamma_1^{(2)} & \gamma_2^{(2)} & \gamma_3^{(2)} \\ \vdots & \vdots & \vdots \\ \gamma_1^{(n_e)} & \gamma_2^{(n_e)} & \gamma_3^{(n_e)} \end{bmatrix} = \begin{bmatrix} \boldsymbol{\gamma}^{(1)} \\ \boldsymbol{\gamma}^{(2)} \\ \vdots \\ \boldsymbol{\gamma}^{(n_e)} \end{bmatrix} \in \mathbb{R}^{n_e \times 3} \tag{22}$$

and let each cluster be represented by its coordinate in the feature space $\boldsymbol{\gamma}^{(i)} = [\gamma_1^{(i)}, \gamma_2^{(i)}, \gamma_3^{(i)}]$, $i = 1, 2, \dots, n_e$. Then, an iterative fuzzy *c*-means clustering algorithm [32, 33] is employed to split the feature space into physical and noise classes. Besides splitting, it can determine the degree of physicalness (DoP) for each cluster. This clustering approach is based on the minimization of the following objective function

$$J_m = \sum_{i=1}^{n_e} \sum_{j=1}^{2} \eta_{ji}^m \|\boldsymbol{\gamma}^{(i)} - v_j\|^2 \qquad (23)$$

in which $\eta_{ji}$ is the degree of membership of $\boldsymbol{\gamma}^{(i)}$ to the $j$th class, (physical or noise), $m$ is the fuzzyness factor (typically $m = 2$) and $v_j$ is the center of the classes.

The clustering algorithm performs the following steps:

1- Randomly initiate the cluster membership values $\eta_{ji}$ within the range $[0, 1]$.
2- Calculate the cluster centers, $v_j = \frac{\sum_{i=1}^{n_e} \eta_{ji}^m \boldsymbol{\gamma}^{(i)}}{\sum_{i=1}^{n_e} \eta_{ji}^m}$, $j = 1,2$.
3- Update $\eta_{ji}$ as $\eta_{ji} = \left( \sum_{q=1}^{2} \left( \frac{\|\boldsymbol{\gamma}^{(i)} - v_j\|}{\|\boldsymbol{\gamma}^{(i)} - v_q\|} \right)^{\frac{2}{m-1}} \right)^{-1}$, $j = 1,2$ and $i = 1,2, \ldots, n_e$.
4- Calculate the objective function $J_m$, in Eq. (23).

Steps 2–4 are repeated until the step-wise reduction of $J_m$ is less than a specified threshold or until a specified maximum number of iterations is reached. Then the factor $0 \leq \eta_{ji} \leq 1$ indicates how much the clusters $c_i$ $i = 1,2, \ldots, n_e$ belong to each class. Considering the physical mode class, this factor $\eta_{ji}$ is the DoP of the cluster $i$. A cluster is thus classified as physical provided that its DoP$\geq$ 0.5, and as noise otherwise.

## 4 Application examples

To provide feasibility evidence of the proposed automatic modal parameter estimation algorithm, two case studies are considered. As a first case, simulation data from a Finite Element (FE) model of an aluminum plate serves as a test case for which there is a known solution. In the second case, real test data collected from a structure struck by a hammer impact are used to give plausible evidence that the proposed method works. For both of them, the data are provided in the form of accelerance FRF. Both cases contain several sets of multiple or almost coalescent modes. Both cases are limited to have two independent excitation forces, and thus the maximum multiplicity of the eigenfrequencies which can be considered is two. In addition to that, the second case contains highly damped modes and marginally controllable modes which are difficult to distinguish from noise modes.

## 4.1  Case study I: aluminum plate with synthetic data

Consider the FE model of a 0.5 m wide square aluminum plate with thickness 1mm as shown in Figure 2, described further in [34]. The plate is simply supported at the four corners. The plate is subjected to two force inputs, $u_1$ and $u_2$ and the acceleration of its out-of-plane motion are captured at eight nodes, $y_i$, $i = 1,2,...,8$ as outputs. FE response data mimics measured accelerance FRFs in the frequency range of $0 - 200$ Hz to give the dataset $\boldsymbol{G}^*$. The data include white Gaussian noise with 5% RMS noise-to-signal ratio added to the noise-free FRFs. The FE model shows 24 modes in this frequency range, among which two of them cannot be observed through this sensor configuration since the accelerometers all sit on nodal lines of these modes. Thus, the target of the proposed algorithm is to estimate modal parameters associated with 22 observable modes (or equally 44 states) present in the frequency range of $0 - 200$ Hz. The motivation for this case study is that the plate has several double modes in the considered frequency range, see Figure 3. This forces clustering-based modal parameter algorithms [7, 19] to use special treatment to separate the multiple modes into distinct clusters.

Step I of the algorithm consists of the identification of an exhaustive model EM using the subspace-based method. Based on a visual inspection of the number of peaks of the accelerance FRFs, see Figure 11, a model order of 200 (100 modes) is considered to be more than sufficient for the EM to capture all physical characteristics present in the frequency range of interest. Therefore, the number of clusters are set to 101 of which 100 are clusters for likely physical modes and one cluster is its complement, the *Trashbox* $\boldsymbol{\mathcal{T}}$, for the noise modes.

In step II, bootstrap samples of the test dataset are taken to provide statistics for the modal properties, such as modal damping, eigenfrequency, and modal contribution. The statistical evaluation is made for the members of the likely physical clusters. To this end, we first search for a model order which is a reasonable overestimation of the true model order by the SVC. Figure 4b demonstrates the SVC criterion values computed based on the singular values of the exhaustive model. As can be seen, the model order 90 gives the minimum SVC value which is higher than the true model order 44. It is worth mentioning that by plotting the modal contribution of the modes of the exhaustive model, as shown in Figure 4a, one can observe a rather significant drop at order 46 which could have been used as a rough estimation of the model order. However, it will be shown by the next case that modal contribution drop cannot always be as decisive as in this example. We generate 100 bootstrap realization datasets to get statistically significant data and identify state-space models of order 90 to each such individual dataset.

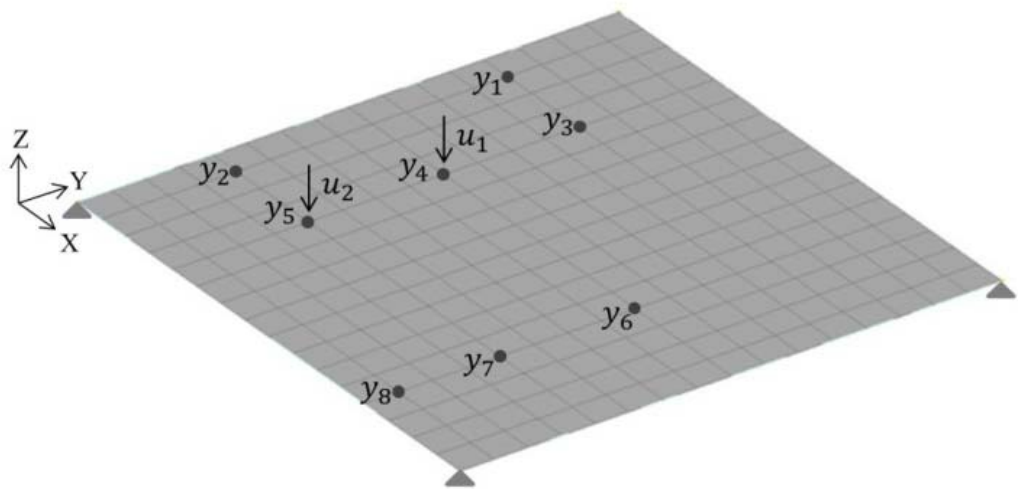

**Figure 2**. Schematic view of the FE model of the aluminum square plate

Mode No. 2, 15.3 Hz

Mode No. 7, 48.7 Hz

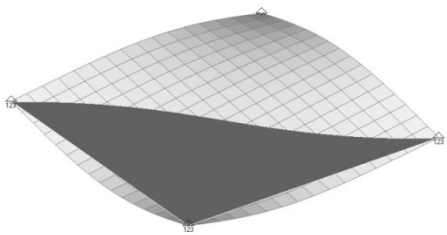
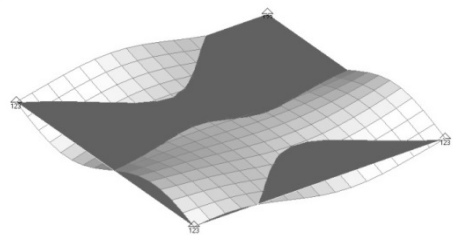

Mode No. 10, 77.7 Hz

Mode No. 13, 110.6 Hz

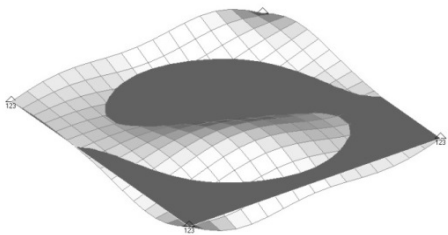
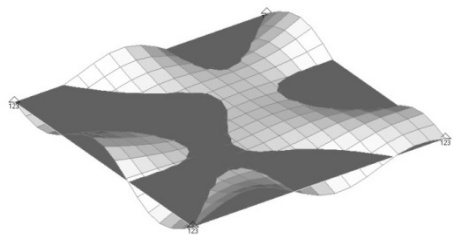

Mode No. 19, 162.6 Hz

Mode No. 22, 190.0 Hz

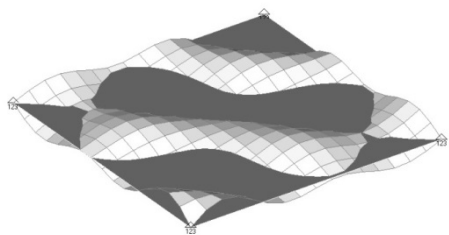
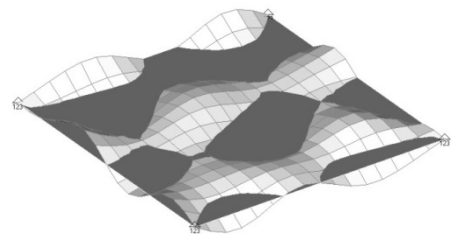

**Figure 3**. One of the mode shape pairs corresponding to double eigenfrequencies of the plate.

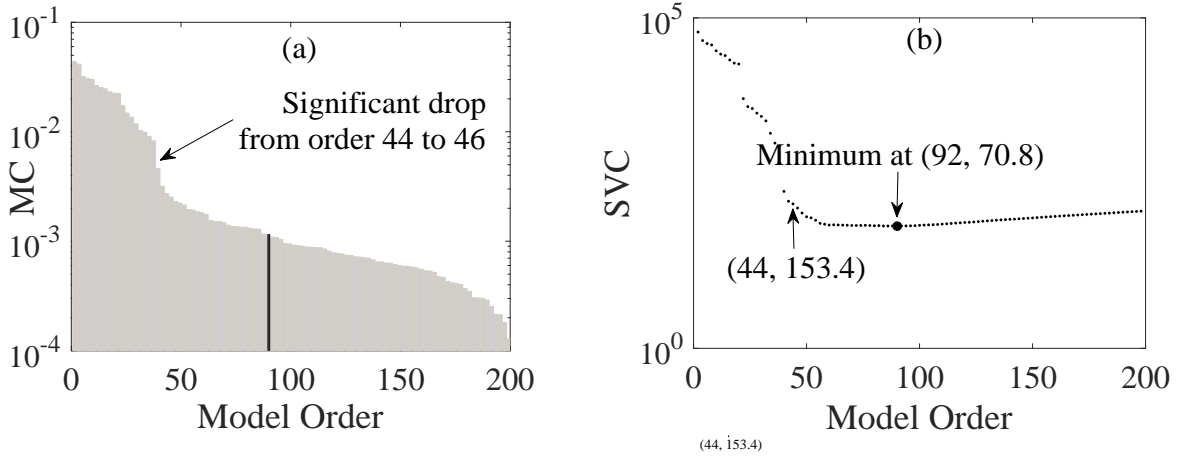

**Figure 4**. Model order selection by (a) SVC. Bullet is for model order 90. (b) Modal contribution only for even model orders. Black column is for model order 90.

In step III of this algorithm, the estimated modes of the bootstrap models are divided into the 101 clusters. This is done using correlation analysis. The correlation of each individual mode with the centroids of the likely physical clusters is measured using the $b$MOC, and the correlation with the noise subspace $\mathcal{N}$ is measured with $h$MOC. Each mode of the BMs is grouped with the cluster centroids to which it shows the highest degree of correlation. The correlation number between modes of two typical BMs and the cluster centroids are shown in Figure 5. The rightmost column, #101, represents subspace $\mathcal{N}$. Modes with highest correlation to this subspace are classified as noise modes and are removed from further analysis. The other modes are classified as likely physical modes and are divided into the 100 likely physical clusters.

In order to illustrate the outcome of the correlation-based clustering algorithm, a correlation analysis is made on all modes collected into the likely physical clusters and their associated centroids, see Figure 6. It shows that the cluster centroids collect similar modes from the pool of modes constructed by the bootstrap models. Yet, there are 7 clusters which remained empty after clustering, see Figure 6. This means their representative modes were likely noise modes realized in the EM. Such clusters are removed and the subspace of likely physical modes are reduced accordingly.

It is worth mentioning that in total there are 3025 modes collected to likely physical clusters. Compared with the number of all $100 \times 45 = 4500$ modes realized in the bootstrapped models, this indicates that the subspace $\mathcal{N}$ collects 1475 modes in the *Trashbox* $\mathcal{T}$.

Figure 7 illustrates the effect of the QR-decomposition for treating multiple modes. It demonstrates the outcome of a correlation analysis between the cluster of likely physical modes with and without using QR-decomposition. Considering the framed clusters belonging to double modes, one notice that excluding QR-decomposition from the algorithm results in a strong correlation between the modes of two associated clusters. This attributes to the fact that every eigenvector which lies in the invariant subspace of a double-mode can be considered as an eigenvector. Thus, the eigenvectors are not unique and may appear differently in each BM. As a result, they give strong but random correlation to the two cluster centroids and a clustering would behave seemingly random. Figure 7b demonstrates that the QR-decomposition fixes the eigenvectors and in effect diminish the correlation of BM modes to one cluster and increases it to the other. The QR-decomposition thus enables a more distinctive separation between the clusters.

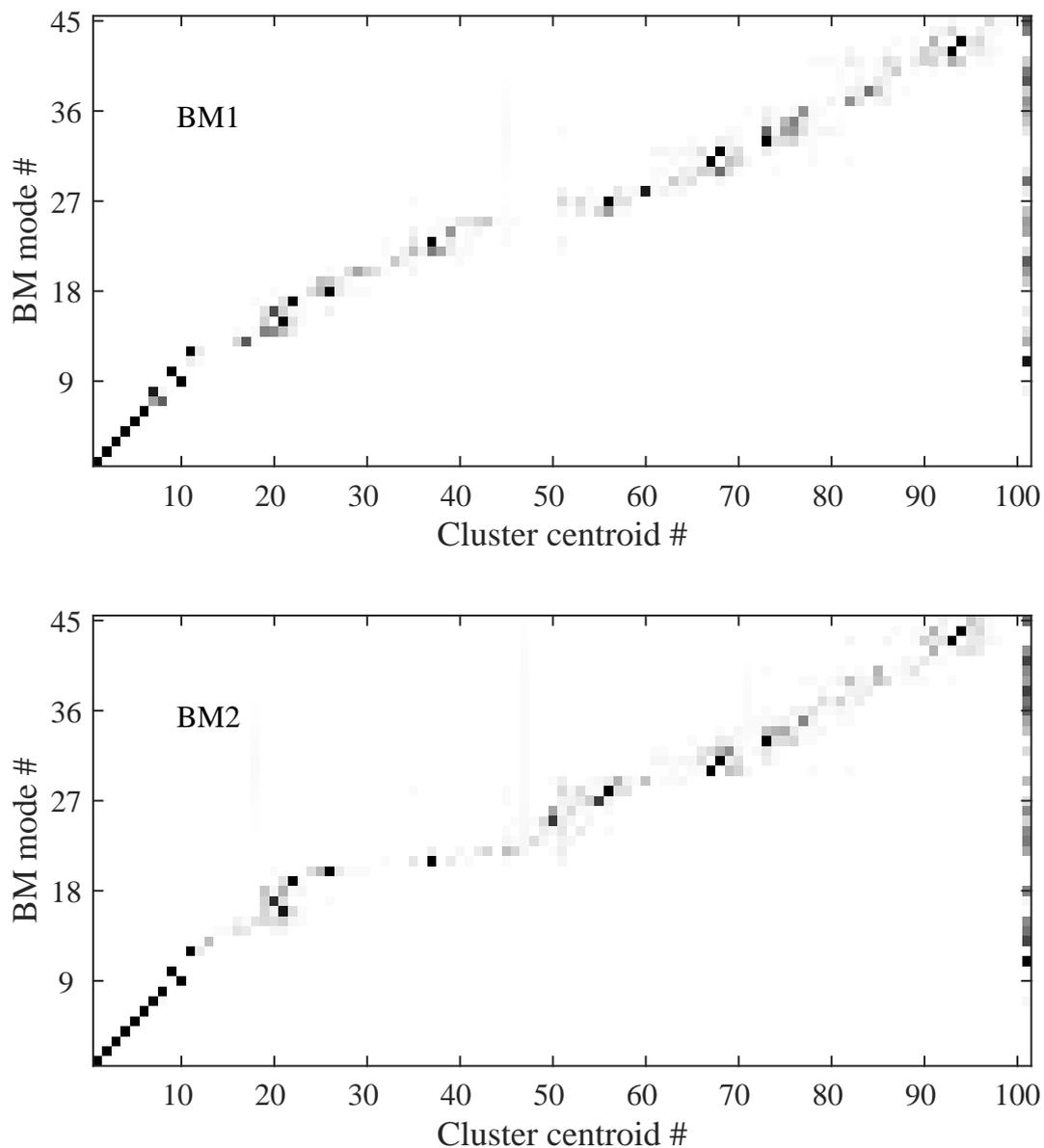

**Figure 5**. Two examples of correlation between bootstrap model modes and cluster centroids. Column 101 is for the correlation with the subspace of the noise modes, $\mathcal{N}$. Grayscale is used in which black represents 100% correlation

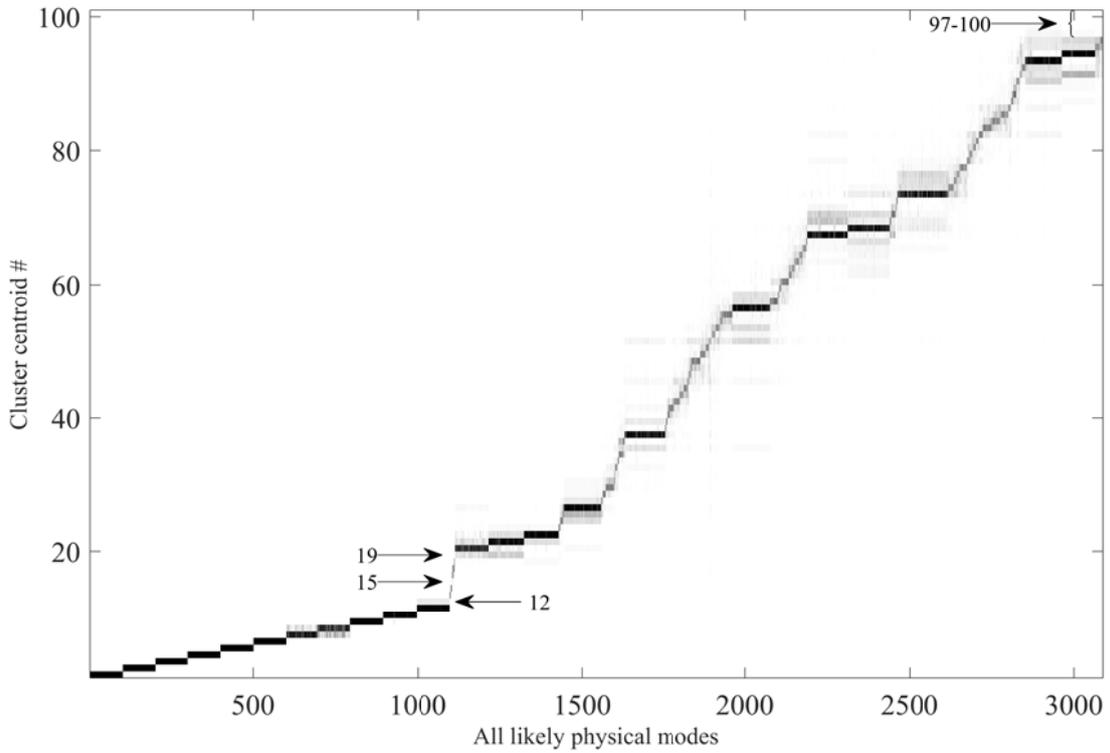

**Figure 6**. Correlation analysis between the members of likely physical clusters and cluster centroids. Empty clusters are indicated by numbered arrows.

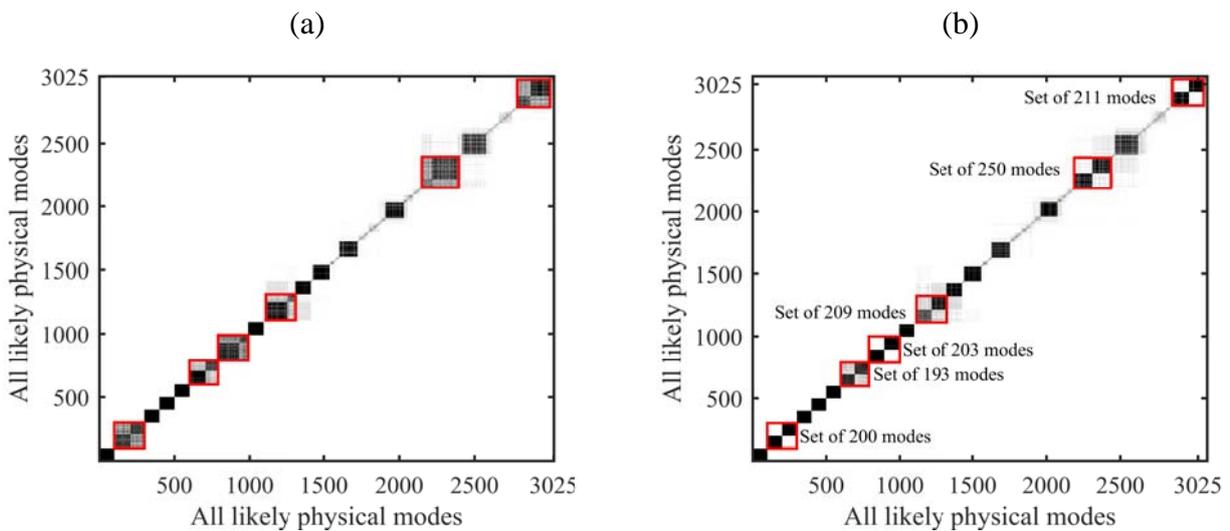

**Figure 7**. Auto-MOC of all likely physical modes before cluster trimming without (a) and with (b) utilizing QR-decomposition.

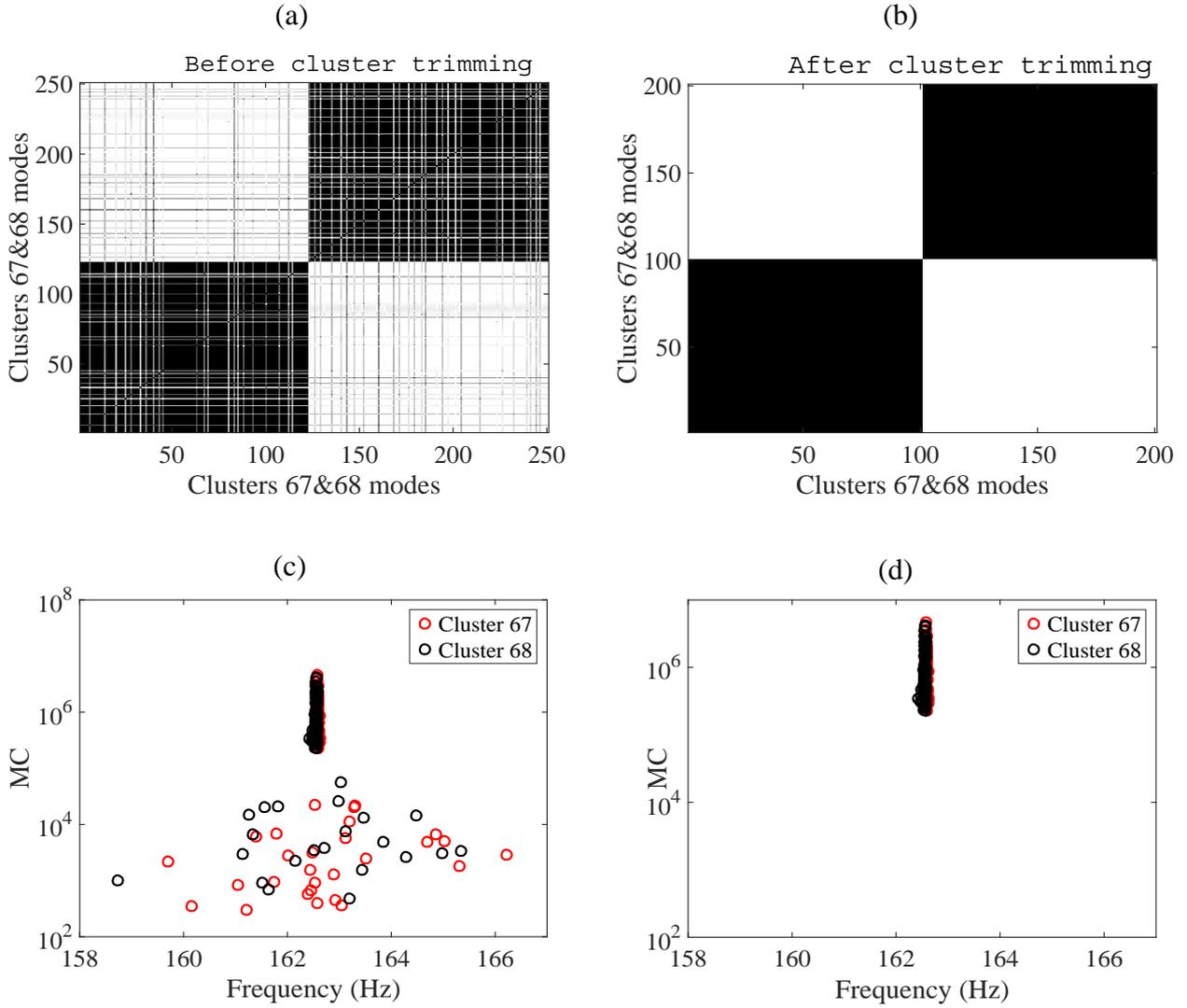

**Figure 8**. Close look at the Auto-MOC and modal contribution of the modes in the clusters associated with 250 eigenfrequencies at 162 Hz before (a,c) and 200 eigenfrequencies after (b,d) trimming.

Step VI of the proposed algorithm consists of cluster trimming. While the *Trashbox 𝒯* has already collected most of the noise modes, there may still exist some noise modes with stronger correlation to the centroids of the likely physical clusters than to the *Trashbox 𝒯*. Yet, they are not highly correlated with the bulk of the other modes of that cluster. For instance, a closer look at clusters #67 and #68 corresponding to the double mode at 162.6 Hz containing 122 and 128 modes is provided by Figure 8a. In this figure the noise modes in each cluster show distinctively lower correlation to the other modes of its cluster. They also show smaller modal contributions as compared to the others, see Figure 8c. These noise modes are identified by the correlation analysis explained in Section 3.4. They are removed from their associated cluster and put in the *Trashbox 𝒯*. For the clusters #67 and #68 this meant that 22 and 28 modes were removed. In addition, each cluster should not contain more modes than the number of bootstrap datasets $N_\mathcal{B}$. In this example the maximum number of modes in a cluster is thus 100. Should it be required, a correlation analysis with the dominant modes of the cluster determines which 100 modes should stay as explained in Section 3.4. The result of such analysis for clusters #67 and #68 is shown in Figure 8b which illustrates two trimmed clusters. The high correlation between modes in a cluster and the similarity between the modal contributions of the remaining modes indicate that the trimming procedure works well.

In step V, three features are extracted for each cluster as shown in Figure 9. The fuzzy c-means clustering with $c = 2$ is employed to assign a degree of physicalness to each cluster. The clustering algorithm terminated successfully with 22 of the remaining clusters having DoP$> 0.5$ and thus indicates correctly that the appropriate model order is 44.

To validate the proposed algorithm in estimating the eigenfrequencies the stabilization diagram is shown in Figure 10 with the Complex Mode Indicator Functions (CMIF) [35] shown in the background. The CMIF confirms the multiplicity of some modes. Vertical lines, located at the average of the eigenfrequency of the modes collected in the clusters, show the DoP of the clusters. For the 6 double modes, a zoom views of the almost coalescent eigenfrequencies are also provided. They indicate that the proposed algorithm captured 22 physical modes from noisy data with good accuracy, even in the presence of repeated eigenvalues.

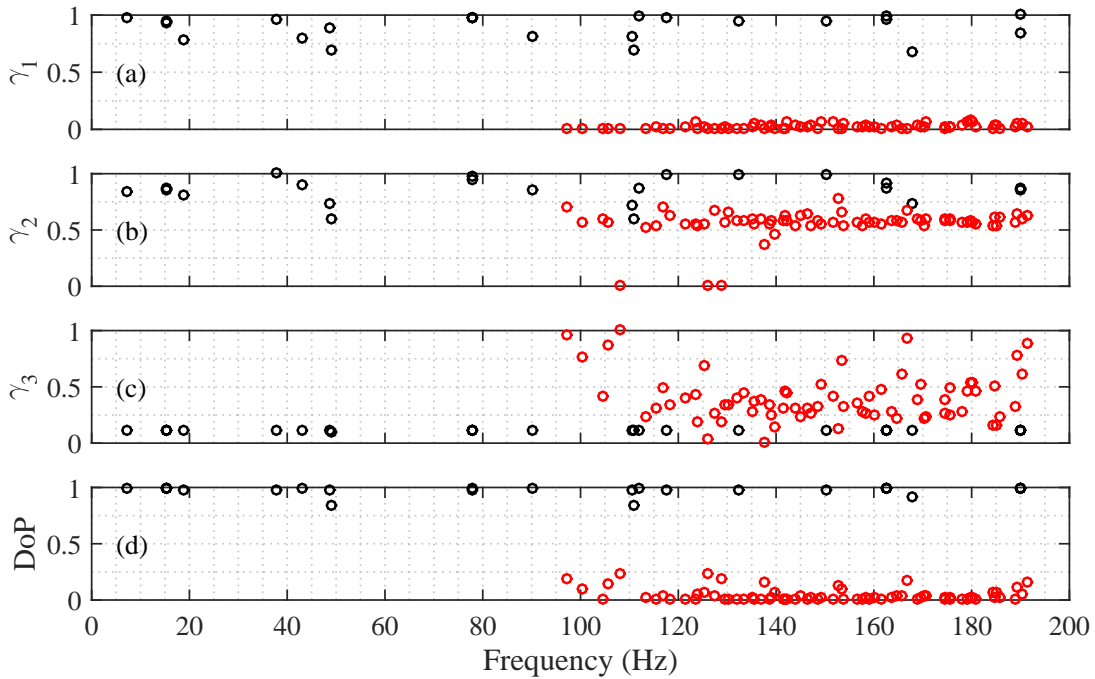

**Figure 9**. (a-c) Feature space of the plate with $k = 1, 2, \dots, 93$. Red is for the noise clusters and black is for physical clusters. (d) Outcome of the fuzzy clustering algorithm.

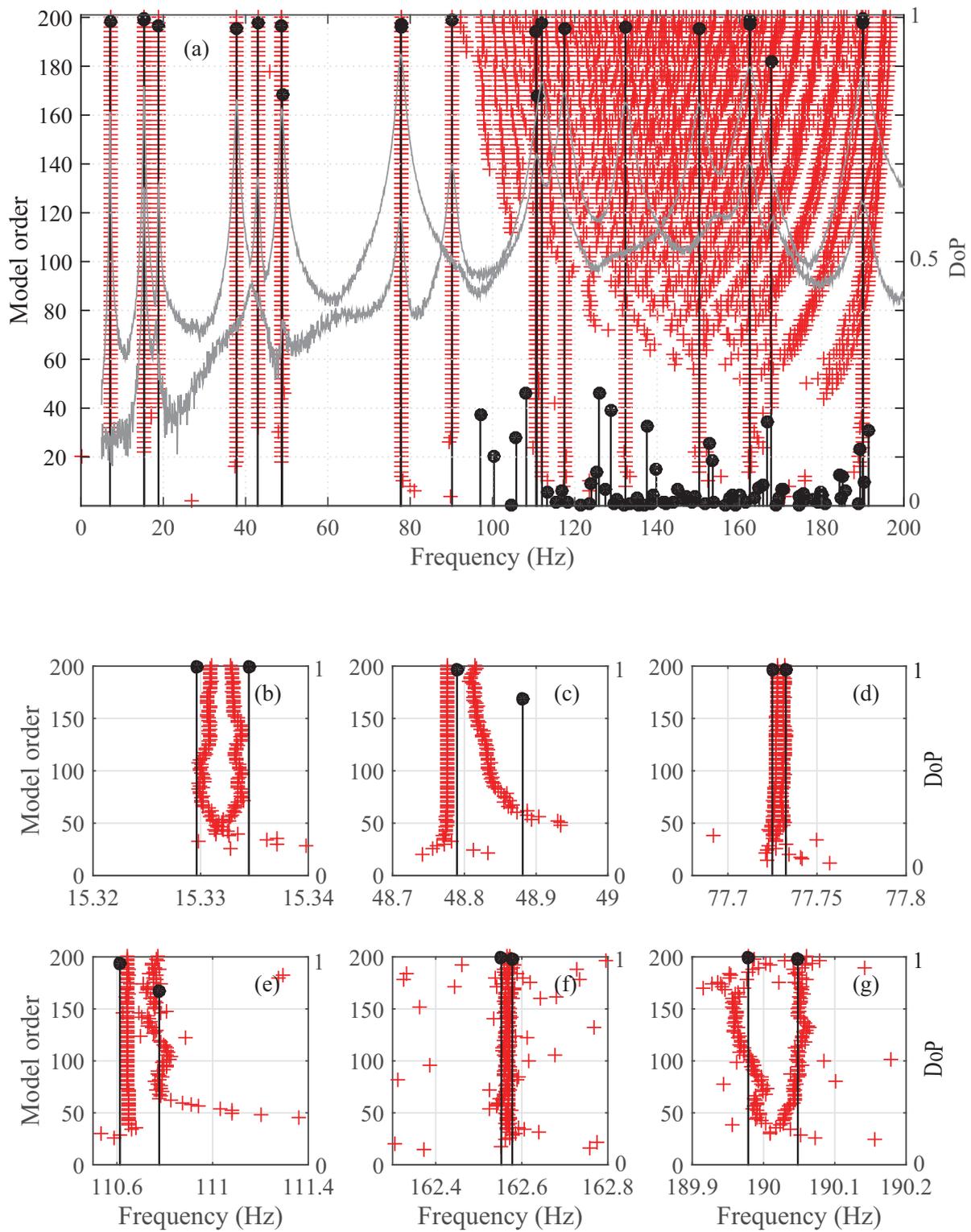

**Figure 10**. (a) Stabilization diagram with the CMIF in grey in the background. Vertical black lines represent the average eigenfrequencies of the modes in the clusters and their height show the DoP of the corresponding modes. (b) through (g) are zoom-in views of all 6 double modes of the plate below 200 Hz.

Figure 11 shows the sum of the magnitude of the transfer functions $|G^*|$ for the two stimuli over all sensors for both input excitations together with the identified physical modes. It shows that the proposed algorithm successfully classifies the physical modes. It is worth noting that an important benefit of employing bootstrapping is the quantification of the uncertainty on the identified modal parameters. Figure 12a demonstrates the spread of the real and imaginary parts of the estimated physical eigenvalues and Figure 12b shows a zoom-in plot for the modes around 111 Hz.

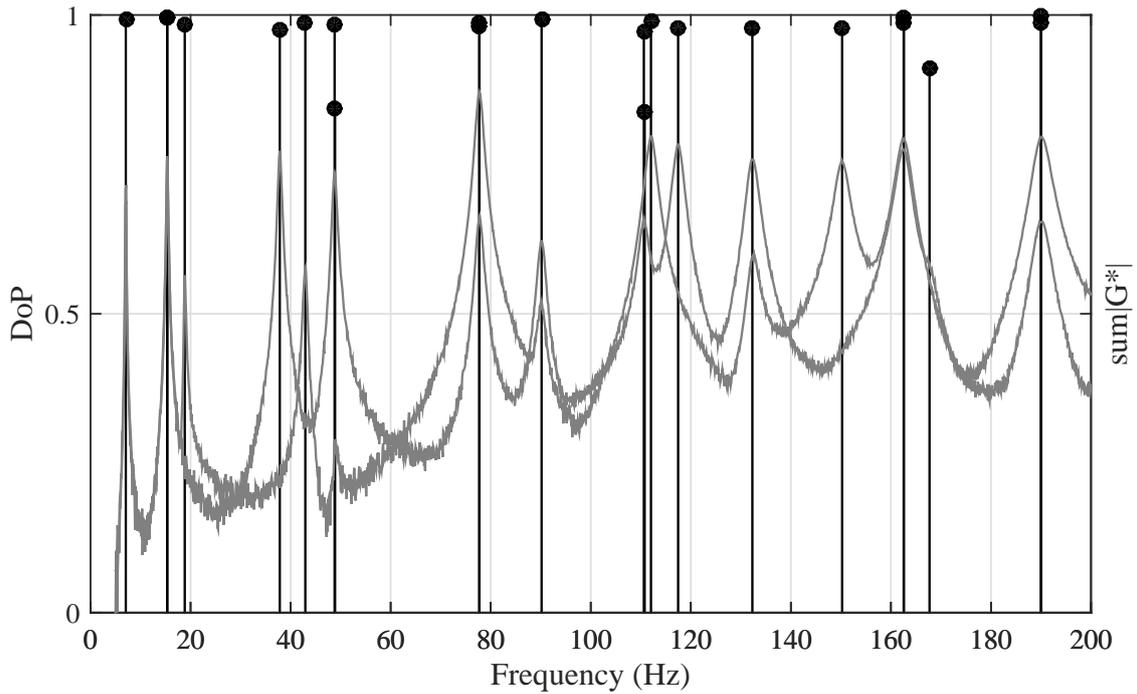

**Figure 11**. Sum of magnitude of transfer functions $|G^*|$ over all sensors and for two input excitations. The vertical black lines show the mean of the eigenfrequency of the identified physical modes. The height of these lines show the DoP of the corresponding modes.

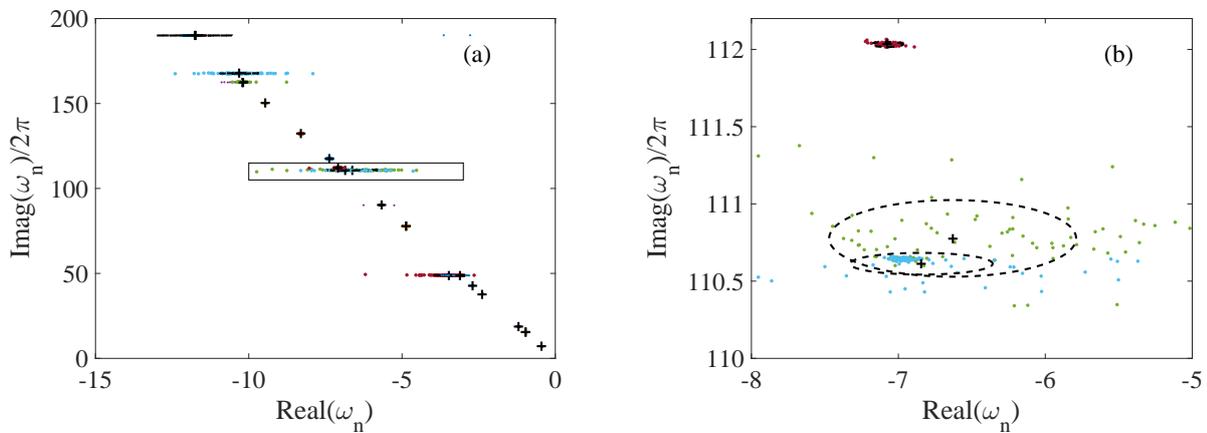

**Figure 12**. (a) Spread of the identified eigenvalues of BM models. (b) Zoom-in of rectangle area at around 111 Hz illustrate one standard deviation boundaries for the real and imaginary parts.

## 4.2 Case study II: pentagon-shaped structure with real test data

In this example, the algorithm is applied to real test data collected from the impact hammer test of a pentagon-shaped steel structure. The pentagon is shown in Figure 13a together with the instrumentation. At two sides of the pentagon, two small identical substructures are connected causing several coalescent modes shown by the frequency responses of the structure. These substructures consist of two parallel plates connected by helical springs. The structure hangs in three long flexible strings at three points, shown by arrows in the figure, which mimic a free-free support. It is subjected to two force inputs, $u_1$ and $u_2$, and its response $y_i$, $i = 1,2,\dots,13$ is measured by uniaxial accelerometers at thirteen nodes. The structure together with the force and sensor locations are shown schematically in Figure 13b.

The data is provided in the form of accelerance FRF in the frequency range [40, 600] Hz. The resolution is 0.05 Hz for the range [40, 200] Hz and 0.1 Hz for the range [200, 600] Hz. Figure 14 shows four of the 26 FRFs. They relate to the inputs $u_1$ and $u_2$ and the accelerations $y_5$ and $y_{13}$.

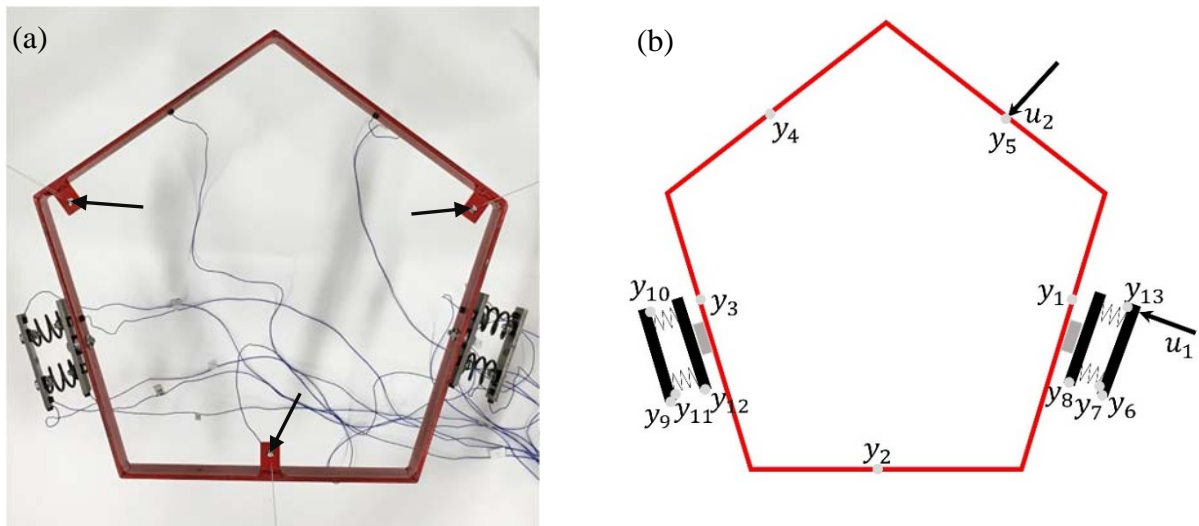

**Figure 13**. a) Top view of the test structure and the instrumentation. Arrows show attachment points for strings used for hanging the frame structure into a horizontal configuration. b) Schematic of the pentagon together with the inputs $u_1$ and $u_2$ and outputs $y_1$ through $y_{13}$.

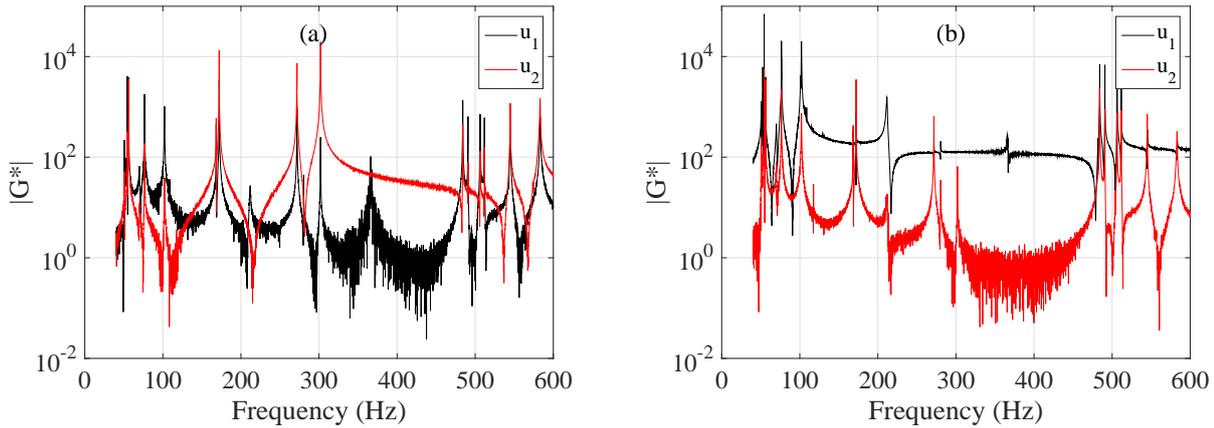

**Figure 14**. Four FRFs of the pentagon between the inputs $u_1$ and $u_2$ to: a) $y_5$, b) $y_{13}$

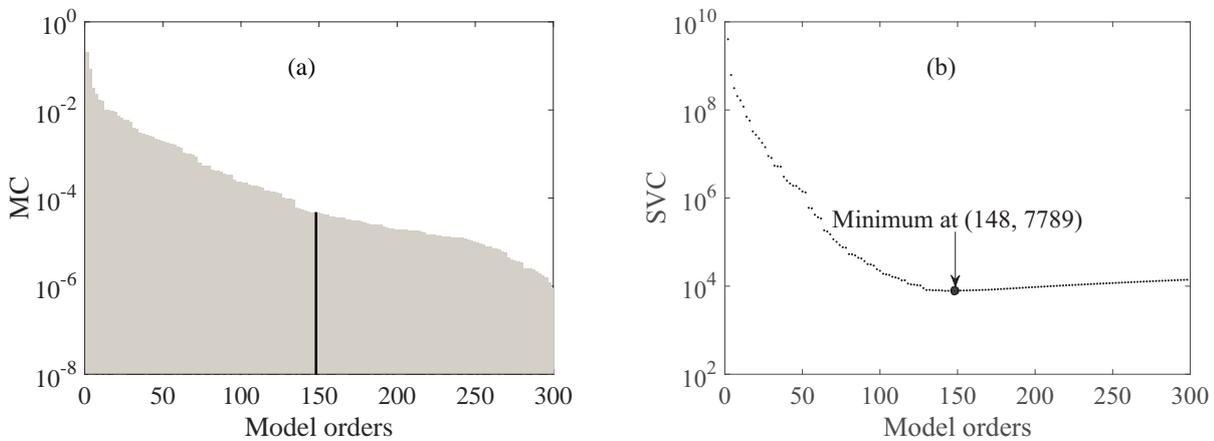

**Figure 15**. Model order selection by the SVC. a) Modal contribution for even model orders. Black column is for model order 148, b) SVC criterion, bullet is for model order 148.

The EM is obtained by identification of the data with a model order 300. This order was deemed high enough by inspection of the collected FRFs. This leads to 151 cluster centroids, 150 for likely physical clusters and one for the *Trashbox 𝒯*.

Figure 15a illustrates the modal contribution to the input-output relation for the EM. The main challenge of this example is the fact that the modal contribution decreases gradually without any obvious large drop. This is due to the presence of highly-damped or marginally controllable physical modes in the frequency range of interest. Therefore, deciding on the number of physical modes only by looking at the modal contribution may lead to either including some noise modes in or excluding some physical modes from the model.

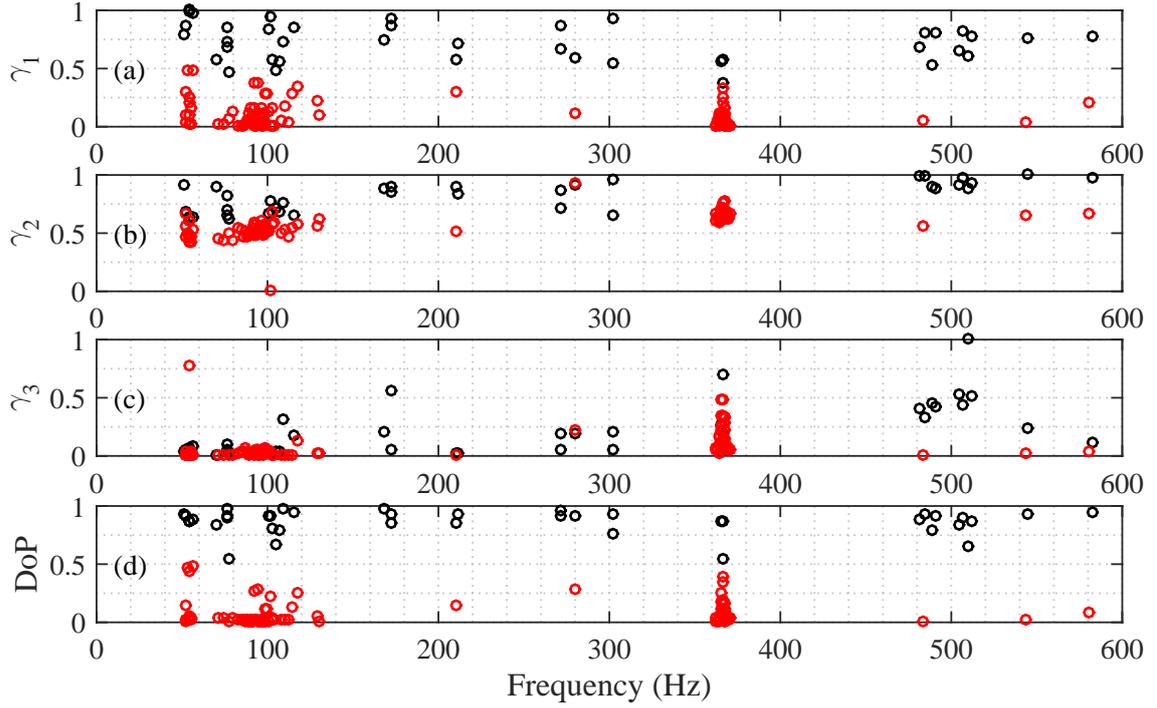

**Figure 16**. (a-c) Feature space of the pentagon structure with $k = 1, 2, \ldots, 150$. Red is for the noise clusters and black is for physical clusters. (d) Outcome of the fuzzy clustering algorithm.

To obtain sufficiently rich statistical data from the BMs, an ensemble of 100 bootstrap datasets are generated. Each of them is identified by a state-space model of order 148. This model order was estimated by the SVC, as was shown in Figure 15b. Then, the modes of the BMs are assigned to the clusters whose representatives give the highest correlation. The cluster trimming step is performed to remove the noise modes from the likely physical clusters by use of correlation metrics $b\text{MOC}$ and $h\text{MOC}$.

The features extracted from the trimmed clusters are shown in Figure 16 (a to c) versus the clusters' eigenfrequency mean. The fuzzy c-means algorithm with $c = 2$ is performed with this feature space to assign a DoP to each cluster. Its outcome is shown in Figure 16d. Physical and noise modes are shown in black and red colors respectively, with 40 physical clusters with DoP$\geq 0.5$. This means that the obtained model order is 80.

For further consideration about the physicalness of the modes, especially the modes with DoP around 0.5, the stabilization diagram including CMIF is provided in Figure 17. The vertical lines are representatives for the clusters and their lengths are proportional to their associated DoPs. Two of the challenging parts of the FRFs with a high density of physical and noise modes are shown in more detail in Figure 17(b and c).

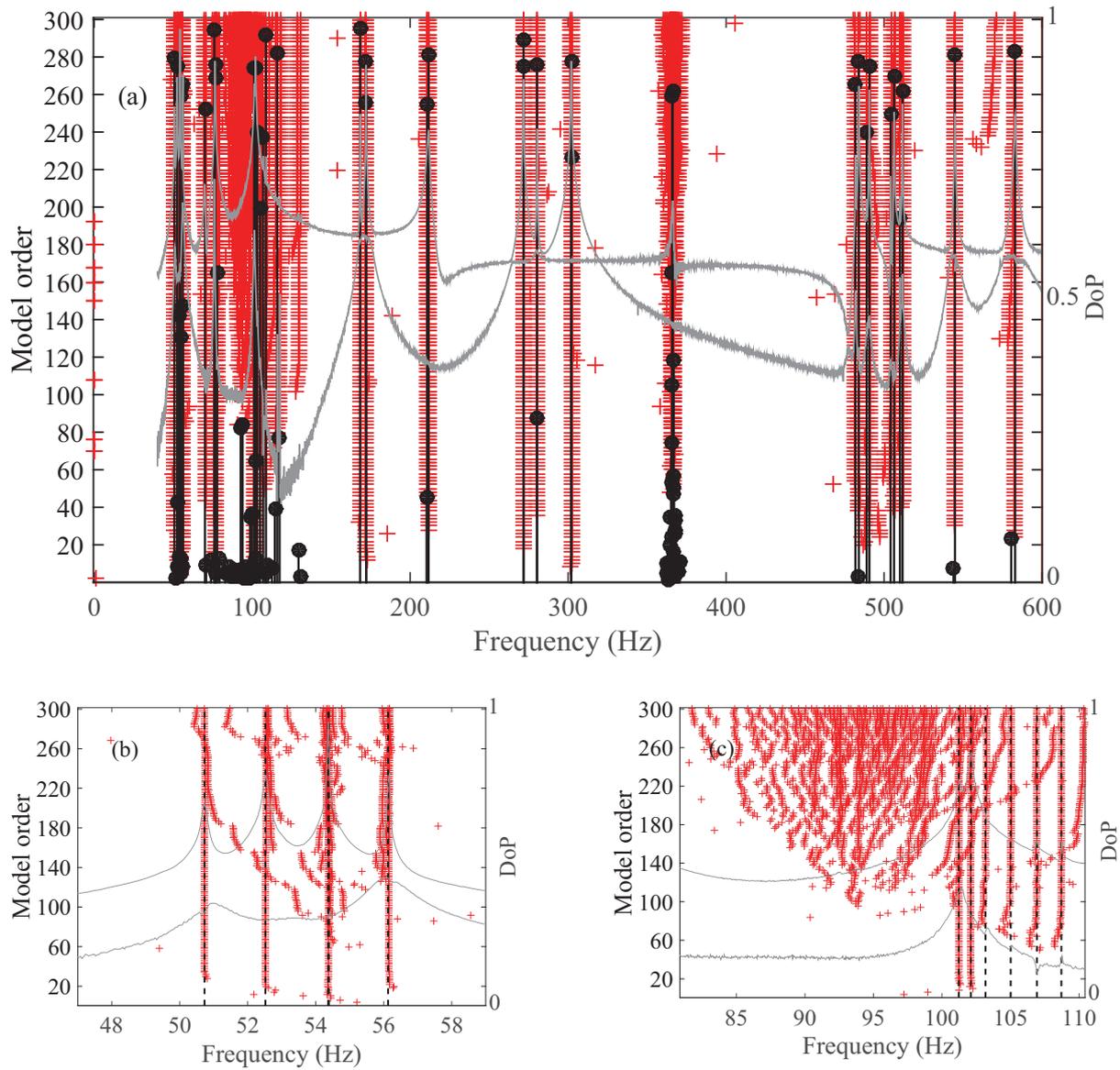

**Figure 17**. a) Stabilization diagram together with CMIF. Vertical lines represent the DoP at the average eigenfrequencies of the modes in the clusters. (b) and (c) show close-up views related to closely spaced modes of the pentagon structure.

It is observed in Figure 16d that there are three modes with DoP a little smaller than 0.5 in the [50, 60] Hz range. A closer look at this region in the stabilization diagram, in Figure 17b, shows that they are at 53.4, 54.7 and 56.2 Hz with DoP values 0.46, 0.43 and 0.49, respectively. As can be seen, there are no stable modes at 53.4 and 54.7 Hz, which indicate that they are noise modes. However, there is a stable mode at 56.2 Hz which starts to appear when the model order is increased above 120. Its appearance at such high model order indicates the negligible contribution of this mode and thus it can be ignored. Figure 17c shows the region around 100 Hz in which many modes appear at high model orders. Among them, several highly-damped or low-excited physical modes between 100 Hz and 110 Hz can be observed. The algorithm distinguished them from the noise modes and captured them as physical modes.

In Figure 18 the summation of the FRFs over the output channels is shown together with DoP indicators of the modes classified as physical. It indicates that the proposed algorithm selects a proper set of physical modes.

The uncertainty bound on the eigenvalues obtained by bootstrap statistics are shown in Figure 19a. The ellipses illustrate one standard deviation bounds of the real and imaginary parts of the eigenvalues. To illustrate the spread better, the zoom-in plot of an area around 54 Hz is provided. To distinguish the coalescent modes at 54 Hz, an even closer zoom-in is also provided.

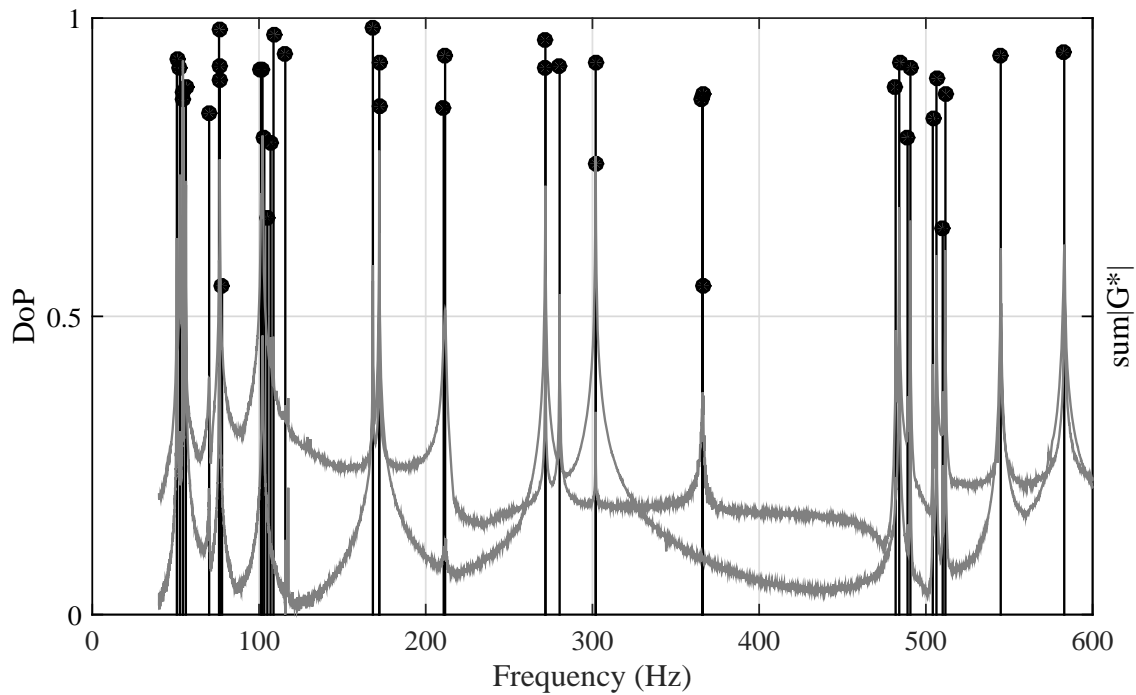

**Figure 18**. Sum $|G^*|$ over the output channels for two inputs. The positions of the black lines indicate the mean of the eigenfrequency of the modes in the clusters classified as physical. The height of these lines shown the DoP of the modes.

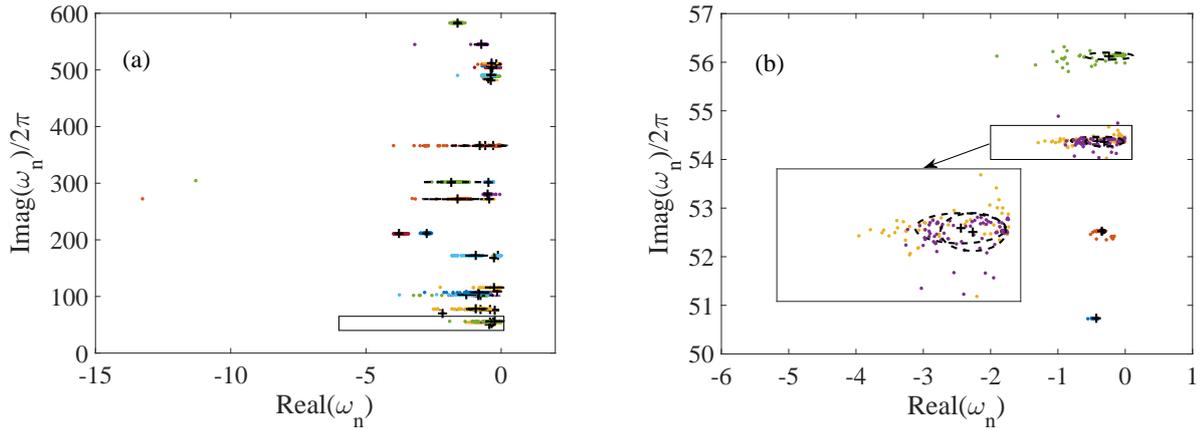

**Figure 19**. (a) Spread of the identified eigenvalues of the structure. (b) Zoom-in of the rectangle area at around 54 Hz illustrate the confidence bounds. A closer zoom-in view is provided for the double modes at around 54 Hz.

## 5  Summary and conclusion

An algorithm for automated modal parameter estimation from noisy data has been developed. It can provide statistical information for the estimated parameters without the need for high dimensional optimization by utilizing a bootstrap sampling in conjunction with a subspace-based identification algorithm. The benefit of a new modal characteristic vector called modal observability vector has been demonstrated. Through analysis and application, it was shown that the correlation analyses which have been developed based on this vector have the capability to deal with the two major challenges in the field of modal correlation, namely (*i*) the spatial aliasing phenomenon and (*ii*) the non-unique eigenvector of the modes of coalescent eigenvalues. Moreover, it was shown that coupling the modal observability vector, correlation analysis, and subspace-based linear algebra provides a platform to cluster the similar modes together and reject noise modes in an automated fashion. In a separate step, all clusters have been scanned for the presence of noise modes which are trimmed from the clusters. A three-dimensional feature space is constructed to which a fuzzy c-means clustering is applied to assign a degree of physicalness to each cluster. The proposed method has been successfully applied to two cases, a case with synthetic test data and a case with real test data. Both indicates the method's adequacy to distinguish between physical and noise modes. These results have been validated by comparing with those obtained using the stabilization chart and a CMIF analysis.

**References**

[1] M. El-Kafafy, T. De Troyer, P. Guillaume, Fast maximum-likelihood identification of modal parameters with uncertainty intervals: A modal model formulation with enhanced residual term, Mechanical Systems and Signal Processing, 48 (2014) 49-66.

[2] M. El-kafafy, T. De Troyer, B. Peeters, P. Guillaume, Fast maximum-likelihood identification of modal parameters with uncertainty intervals: A modal model-based formulation, Mechanical Systems and Signal Processing, 37 (2013) 422-439.

[3] H. Akaike, A new look at the statistical model identification, IEEE Transactions on Automatic Control, 19 (1974) 716-723.

[4] D. Bauer, Order estimation for subspace methods, Automatica, 37 (2001) 1561-1573.

[5] T. Saito, J.L. Beck, Bayesian model selection for ARX models and its application to structural health monitoring, Earthquake Engineering & Structural Dynamics, 39 (2010) 1737-1759.

[6] S. Vanlanduit, P. Verboven, P. Guillaume, J. Schoukens, An automatic frequency domain modal parameter estimation algorithm, Journal of Sound and Vibration, 265 (2003) 647-661.

[7] E. Reynders, J. Houbrechts, G. De Roeck, Fully automated (operational) modal analysis, Mechanical Systems and Signal Processing, 29 (2012) 228-250.

[8] P. Verboven, E. Parloo, P. Guillaume, M. Van Overmeire, Autonomous structural health monitoring - Part I: Modal parameter estimation and tracking, Mechanical Systems and Signal Processing, 16 (2002) 637-657.

[9] P. Verboven, E. Parloo, P. Guillaume, M. Van Overmeire, Autonomous modal parameter estimation based on a statistical frequency domain maximum likelihood approach, IMAC XIX, Florida, USA, 2001.

[10] V. Yaghoubi, T. Abrahamsson, Automated modal analysis based on frequency response function estimates, Topics in Modal Analysis I, Volume 5, Springer, 2012, pp. 9-18.

[11] I. Goethals, B. De Moor, Model reduction and energy analysis as a tool to detect spurious modes, in: Proc. International Conference on Noise and Vibration Engineering (ISMA), Leuven, Belgium, 2002, pp. 1307-1314.

[12] I. Goethals, B. Vanluyten, B. De Moor, Reliable spurious mode rejection using self learning algorithms, in: Proc. International Conference on Noise and Vibration Engineering (ISMA), Leuven, Belgium, 2004, pp. 991-1003.

[13] F. Nasser, Z. Li, P. Gueguen, N. Martin, Frequency and damping ratio assessment of high-rise buildings using an automatic model-based approach applied to real-world ambient vibration recordings, Mechanical Systems and Signal Processing, (2016).

[14] F. Nasser, Z. Li, N. Martin, P. Gueguen, An automatic approach towards modal parameter estimation for high-rise buildings of multicomponent signals under ambient excitations via filter-free Random Decrement Technique, Mechanical Systems and Signal Processing, 70–71 (2016) 821-831.

[15] C. Rainieri, G. Fabbrocino, Automated output-only dynamic identification of civil engineering structures, Mechanical Systems and Signal Processing, 24 (2010) 678-695.

[16] S. Qin, J. Kang, Q. Wang, Operational modal analysis based on subspace algorithm with an improved stabilization diagram method, Shock and Vibration, (2016).

[17] C. Rainieri, G. Fabbrocino, Development and validation of an automated operational modal analysis algorithm for vibration-based monitoring and tensile load estimation, Mechanical Systems and Signal Processing, 60–61 (2015) 512-534.

[18] F. Magalhães, Á. Cunha, E. Caetano, Online automatic identification of the modal parameters of a long span arch bridge, Mechanical Systems and Signal Processing, 23 (2009) 316-329.

[19] A.W. Phillips, R.J. Allemang, D.L. Brown, Autonomous modal parameter estimation: Methodology, in: Proc. IMAC XXVIII, Florida, USA, 2011, pp. 363-384.

[20] M. Scionti, J. Lanslots, Stabilisation diagrams: Pole identification using fuzzy clustering techniques, Advances in Engineering Software, 36 (2005) 768-779.

[21] T. McKelvey, H. Akçay, L. Ljung, Subspace-based multivariable system identification from frequency response data, IEEE Transactions on Automatic Control, 41 (1996) 960-979.

[22] R.J. Allemang, The modal assurance criterion - Twenty years of use and abuse, Sound and Vibration, 37 (2003) 14-21.

[23] V. Yaghoubi, T. Abrahamsson, The modal observability correlation as a modal correlation metric, in: Proc. IMAC XXXI, California, USA, 2013, pp. 487-494.

[24] A.W. Phillips, R.J. Allemang, Data presentation schemes for selection and identification of modal parameters, IMAC XXIII, Florida, USA, 2005.

[25] A.J. Laub, Matrix Analysis For Scientists And Engineers, Society for Industrial and Applied Mathematics, 2004.

[26] B. Efron, Bootstrap methods: another look at the jackknife, The Annals of Statistics, (1979) 1-26.

[27] B. Efron, R.J. Tibshirani, An introduction to the bootstrap, CRC press, 1994.

[28] M. Khorsand Vakilzadeh, S. Rahrovani, T. Abrahamsson, An improved modal approach for model reduction based on input-output relation, in: Proc. International Conference on Noise and Vibration Engineering (ISMA), Leuven, Belgium, 2012, pp. 3451-3459.

[29] R.J. Allemang, A.W. Phillips, D.L. Brown, Autonomous modal parameter estimation: Statisical considerations, in: Proc. IMAC XXIX, New York, USA, 2011, pp. 385-401.

[30] I. Goethals, B. De Moor, Subspace identification combined with new mode selection techniques for modal analysis of an airplane, in: Proc. 13th IFAC symposium on system identification (SYSID 2003), Rotterdam, Netherlands, 2003, pp. 695-700.

[31] L. Kaufman, P.J. Rousseeuw, Finding groups in data: an introduction to cluster analysis, John Wiley & Sons, 2009.

[32] J.C. Bezdek, Pattern recognition with fuzzy objective function algorithms, Springer Science & Business Media, 2013.

[33] I. MathWorks, W.-c. Wang, Fuzzy Logic Toolbox: for Use with MATLAB: User's Guide, Mathworks, Incorporated, 1998.

[34] M.K. Vakilzadeh, S. Rahrovani, T. Abrahamsson, Modal reduction based on accurate input-output relation preservation, in: Proc. IMAC XXXI, Florida, USA, 2014, pp. 333-342.

[35] D.J. Ewins, Modal testing: theory and practice, Research studies press Letchworth, 1984.